\documentclass [a4paper,aip,jmp,preprint,numerical,showkeys,showpacs,unsortedaddress] {revtex4-1}

\makeatletter
\usepackage [english] {babel}
\usepackage [latin1] {inputenc}
\usepackage [T1] {fontenc}

\usepackage {times}
\usepackage {amsmath, amssymb, amsthm, mathrsfs, slashed}
\usepackage {hyperref}

\begin {document}

\title {Charged States In A Semiclassical Model Of Infrared Gupta-Bleuler Quantum Electrodynamics}

\author {Simone Zerella}
\email {simone.zerella@gmail.com}
\affiliation {Dip. di Fisica, Universit\`a di Pisa, Pisa, Italy}

\begin {abstract}
We address the problem of the identification and characterization of charged states within local and covariant 
quantizations of abelian gauge theories, focusing on a semiclassical model of infrared Gupta-Bleuler Quantum 
Electrodynamics, based on the Bloch-Nordsieck approximation and formulated in Feynman's gauge. 
The $\, GNS\, $ construction over suitable functionals yields positive subspaces of the indefinite-metric space of 
the model; charged states with Li\'enard-Wiechert space-like asymptotics can then be constructed via an 
automorphism of the algebra of observables implementing Gauss' law.
Finally, by an analysis of the localization properties of the corresponding expectations over the asymptotic 
electromagnetic fields, it is shown that such states identify an infrared-minimal charge class in the sense 
of Buchholz.

\end {abstract}
\maketitle

\section* {Introduction}

One of the main open questions in the theoretical understanding of Quantum Electrodynamics $(QED)$ concerns 
the construction and the characterization of physical charged states. 
This issue is much more difficult, with respect to a standard Quantum Field Theory $(QFT)\, ,\, $ owing to the 
presence of a local Gauss law.
The latter relates the total charge in a bounded region to the flux of electric field through the boundary 
of the region and by locality of observables can be shown to imply, within a quantum mechanical 
setting, the non locality of the electrically charged states\cite {FePSt74FePSt77} and the 
superselection of electric charge\cite {StWight}. 
By the same token, it implies that the state space of $\, QED\, $ contains uncountably many superselection 
sectors, labeled by the value of the electric flux in suitably chosen spacelike cones\cite {Buch82,Buch86}, 
non-invariant under Lorentz boosts\cite {FrMoSt79a,FrMoSt79b,Buch86} and with single-particle 
subspaces not containing proper eigenstates of the mass operator\cite {Buch86}.

Besides being a fundamental issue in the understanding of the structural properties of (abelian) gauge theories, 
the identification of physical charged states is also necessary for the solution of the infrared problem, namely for 
the characterization of states of charged particles and of the quantized e.m. field at asymptotic times
and for the definition of a scattering matrix. 
In particular, the question arises whether it is possible to formulate a collision theory in terms of an appropriate 
subset of charged states; in this respect, it is plausible that the discussion of the infrared problem may be made 
simpler by choosing the (charged) sectors with the best localization properties relative to the vacuum.

Quite generally, whatever formulation one may employ, the construction of the charged sectors of $\, QED\, $ 
always presents subtle features.
Within the Haag-Kastler formulation\cite {Haagb}, based on nets of algebras of local observables, charged 
sectors should in principle be obtained acting on the vacuum sector with suitable morphisms of the 
observable algebra, by a generalization of the construction of Doplicher, Haag and Roberts\cite {DHRI}; 
however, since the needed morphism cannot be local, by virtue of Gauss' constraint, such a generalization 
is difficult to be realized and has not yet been accomplished.

Wightman's formulation of $\, QFT\, $\cite {StreatWight}, although less economical from a conceptual point 
of view, is closer to the perturbative-theoretic framework and hence can also be used to investigate the 
mathematical structures at the basis of local gauge quantizations and to provide support to the 
perturbative methods.
However, one has to face the problem that quantum gauge theories do not fit completely in such a 
formulation because of the presence of the Gauss law.
 
In fact, quantizations in ``renormalizable'' (local and covariant) gauges, based on unobservable fields, the 
Dirac field and the electromagnetic four-vector potential, necessarily yield an indefinite metric space 
$\, \mathscr {G}\, $ of local states, containing vectors with no physical (quantum-mechanical) interpretation\cite {St67}. 
An additional constraint is then imposed in order to select physical states, following the Gupta-Bleuler $(\, GB\, )$
formulation\cite {Gupt50Bleul50}, and one has to face the problem that states with non-zero electric 
charge cannot exist in $\, \mathscr {G}\, $. 

Quantization of electrodynamics in a gauge only involving physical degrees of freedom, as for instance the 
Coulomb gauge, imply on the other hand that the gauge fields cannot be neither local nor covariant and it 
is therefore difficult to set up a perturbative expansion and the renormalization procedure.

In Reference~\onlinecite {MoSt83MoSt84} it was shown that a possible strategy to determine physical 
charged states in the $\, GB\, $ formulation is to obtain them as weak limits of local states, with the help 
of auxiliary topologies, introduced on $\, \mathscr {G}\, $ in order to obtain weakly complete spaces of local states. 

A different approach, pioneered by Dirac\cite {Dirac55} and subsequently developed by Symanzik\cite {Sym71}, 
is based on the (perturbative) construction of Coulomb-type fields, which yield physical states when applied
to the ground state.

The main problem of this strategy is that in order to give a meaning to the formal exponent, a control of both 
ultraviolet and infrared problems is required.
Such issues have been extensively discussed by Steinmann\cite {Steinb}, who proved the existence of physical 
charged fields for a large class of gauge-fixing functions within the framework of perturbation theory.

The characterization of the properties of charged states was discussed in Ref.~\onlinecite {BDMRS01}, partially 
relying on Steinmann's results; in this respect, it was proven that quantum effects imply non locality properties 
not expected from classical considerations, at least for the states obtained by applying suitably regularized 
exponentials to the vacuum. 

The use of specific physical fields yields electrically charged states whose connection with the local and covariant 
formulation of $\, QED\, $ is rather indirect. 
As a matter of fact, it demands to employ a non-trivial generalization of Feynman's rules in the calculation of 
transition amplitudes and thus to take into account a number of additional contributions, with respect 
to the Feynman-Dyson perturbative theory, whose relevance for the formulation of scattering and for 
the outcome of practical computations is at present unclear.
An example showing the non-triviality of this problem is Steinmann's evaluation of the second-order radiative 
corrections to the magnetic moment of the electron; the proof that the result agrees with the value from 
standard perturbation theory involves cancellations of a vast number of contributions\cite {Stein03}.

The results of the most recent investigations thus motivate on the one hand the need for a better understanding 
of the local and covariant formulation of $\, QED\, ;\, $ on the other hand, they lead to a problem of minimality both in the classification of the charged 
states and in the use of physical charged fields. 

In this paper we consider a hamiltonian (infrared) model in the Feynman-Gupta-Bleuler $(FGB)$ gauge, based 
on an expansion of four-momenta whose relevance for the analysis of the infrared problem was first pointed 
out by Bloch and Nordsieck\cite {BlNor}.

In a previous work\cite {Simone1}, this model, which will be henceforth referred to as $\, BN\, $ model, has been 
shown to allow to fully retrieve the results of the diagrammatic treatment of the infrared contributions of 
$\, QED\, ,\, $ in terms of the expansion of M\"{o}ller operators obtained with the aid of an 
infrared cutoff, a mass renormalization and an adiabatic switching of the interaction.

The present paper is devoted to the study of the space-time properties of the $\, BN\, $ model which can be 
inferred from the solution of the Heisenberg equations and to the construction and the characterization of 
(a class of) physical charged states\cite {Simone}.

First, we show that the $\, GNS\, $ theorem over suitable product functionals $\, \omega_{\,\, G}^{}\, ,\, $ 
corresponding to vectors of the indefinite-metric Gupta-Bleuler space of the model, yields positive 
subspaces, a fact that will be important for the subsequent identification of physical charged 
states.
Secondly, we establish the existence of Haag-Ruelle asymptotic limits for the four-vector potential.
We find then that the Fock property, established\cite {Buch77} for the representations of asymptotic 
algebras corresponding to massless bosons and associated to certain regions of Minkowsky space, in 
models of field theories in which locality and positivity are satisfied, is also enjoyed by the 
representations of the photon asymptotic algebras associated to the same regions and 
induced by $\, \omega_{\; G}^{}\, $, which fulfill locality but \emph {not} positivity.

Concerning the construction of electrically charged states, two possible procedures will be outlined,
both based on the realization of the Gauss law constraint via an automorphism of the observable 
algebra.
The first procedure closely follows the strategy employed in classical electrodynamics and is based upon 
the existence of a solution of the free wave equation with support causally disjoint from that of the 
charge density; the second one is based on the introduction of a (suitably regularized) Dirac-type 
factor, constructed with the aid of the asymptotic electromagnetic fields.

As we shall see, since the requirement of a non-zero electric charge does not fix the automorphism uniquely,
we obtain physical charged states with different large-distance behaviour, indexed by a superselected 
parameter related to the corresponding Li\'enard-Wiechert $\, (L\, W)\, $ potentials.

Moreover, we show that such states define a charge class (a concept introduced by Buchholz in 
Ref.~\onlinecite {Buch82} within the algebraic setting), uniquely associated to the $\, GB\, $ 
formulation.
We also determine the properties of the representations of the asymptotic e.m. field algebras associated 
to suitably chosen lightcones and prove that the states of the $\, GB\, $ charge class fulfill a notion of 
minimality, in the sense that they ``only contain the photons associated to the asymptotic 
momentum of the particle''.

The plan of the manuscript is as follows. 

In the first Section we establish notations, introduce the model and evaluate the solutions of the 
Heisenberg equations for the four-vector potential and the corresponding asymptotic e.m. fields.

In Section 2 we examine the localization properties of the expectations of product functionals over 
the four-vector potential and the asymptotic e.m. fields.

Section 3 is devoted to the construction and to the analysis of the properties of physical charged 
states with Li\'enard-Wiechert space-like asymptotics.
First, we outline the Gupta-Bleuler strategy for the determination of solutions of Maxwell's equations 
in classical electrodynamics and a discuss a semiclassical argument concerning the classification of 
charged states.
Afterwards, physical states are identified and their space-time features are investigated.
We conclude the Section with a discussion regarding questions left open by our treatment, concerning the 
description of particles carrying an electric charge at asymptotic times and the vacuum-polarization 
effects.

\section {Notations And Main Features Of The Model}

In this paper we will make use of the following notations.
\newline

The metric $\, g^{\; \mu\, \nu}=\, diag\,\, (\, 1\, ,\, -\, 1\, ,\, -\, 1\, ,\, -\, 1\, )\, $ of Minkowski space is adopted and 
natural units are used $(\, \hbar\, =\, c\, =\, 1\, )\, .\, $
A four-vector is indicated with $\, v^{\, \mu}\, $ or simply with $\, v\, ,\, $ while the symbol $\, \mathbf {v}\, $ 
denotes a three-vector; when confusion may arise, the notation $\, \underline {\, v}\, $ will be employed.
We use the symbol $\, c \cdot d\, $ for the indefinite inner product between four-vectors $\, c\, $ and $\, d\, . $   

The symbol $\, A^{\; \dagger}\, $ stands for the hermitian conjugate of an operator $\, A\, ,\, $ defined on an 
indefinite-metric space, with respect to the indefinite inner product 
$\, \langle\, .\, ,.\, \rangle\, .$\\
The commutator between two operators will be indicated by $\, [\, .\, , .\, ]\, .$ 
We denote by $\, a^{\, \mu\, }(\, \mathbf {k}\, )\, $ and $\, a^{\, \mu\; \dagger}\, (\, \mathbf {k}\, )\, $ respectively the 
annihilation and creation operator-valued distributions in the $\, FGB\, $ gauge, fulfilling the $\, CCR\, $ 
\begin {equation}
[\,\, a^{\, \mu}\, (\, \mathbf {k}\, )\: ,\; a^{\: \nu\; \dagger\, }(\, \mathbf {k}\, '\, )\; ]\, =
\, -\; g^{\: \mu\, \nu}\,\,\, \delta\; (\, \mathbf {k}-\mathbf {k}\; '\, )\, .
\nonumber
\end {equation}
In the same gauge, the Hamiltonian of the free e.m. field is 
\begin {equation}
H_{\,\, 0}^{\, \,e.\, m.}\, =\, -\int\; d^{\,\, 3\, }k\;\; \vert\, \mathbf {k}\, \vert\;\, 
a^{\: \mu\: \dagger}\, (\, \mathbf {k}\, )\;\,\, a_{\; \mu}^{}\, (\, \mathbf {k}\, )
\nonumber
\end {equation}
and the four-vector potential and the e.m. field tensor at time $\, t=0\, $ are respectively
\begin {equation}  
A^{\; \mu\; }(\, \mathbf {x}\, )\, \equiv\, \int\,\, \frac {\, d^{\,\, 3\, }k} {\sqrt {\; 2\;\, \vert\, \mathbf {k}\, \vert}}
\;\;\, [\,\, a^{\, \mu}\, (\, \mathbf {k}\, )\;\,\, e^{\; i\,\, \mathbf {k}\, \cdot\; \mathbf {x}}\, +\, a^{\: \mu\; \dagger}\, 
(\, \mathbf {k}\, )\;\,\, e^{\, -\, i\,\, \mathbf {k}\, \cdot\; \mathbf {x}}\;\, ]\; ,
\nonumber
\end {equation} 
\begin {equation}  
F^{\; \mu\, \nu\, \,}(\, \mathbf {x}\, )\, \equiv\;\, \partial^{\,\, \mu}\,\, A^{\,\, \nu\; }(\, \mathbf {x}\, )-\,
\partial^{\,\, \nu}\,\, A^{\,\, \mu\; }(\, \mathbf {x}\, )\: .
\quad\quad\quad\quad\quad\quad\quad\quad\quad\quad\;\;\nonumber
\end {equation}
The symbols $\, \Psi_{\; F}\, ,\, \omega_{\; F}^{}\, ,\, \pi_{\; F}^{}\, $ will be used respectively for the no-particle 
vector of $\, \mathscr {F}\, ,\, $ the Fock vacuum functional and the Fock representation.
The convolution with a form factor $\, \rho\, $ is indicated by
\begin {equation}\label {convoluzione covariante}
A^{\; \mu}\; (\, \rho\, ,\, \mathbf {x}\, )\, \equiv\; \int\,\, d^{\,\, 3\, }\xi\;\,\,\, 
\rho\; (\, \mathbf {\xi}\, )\;\,\, A^{\; \mu\; }(\, \mathbf {x}-\, \mathbf {\xi}\; )
\nonumber
\end {equation}
and for brevity we write
\begin {equation}
a\, (f\, (t))\, \equiv\; \int\,\, d^{\,\, 3\, }k\;\,\, a^{\: \mu\; }(\, \mathbf {k}\, )
\,\,\, f_{\; \mu}^{}\: (\, \mathbf {k}\, ,\: t\, )\: .
\nonumber
\end {equation}
$\, \mathscr {S\; }(\, \mathbb {R}^{\, 3}\, )\, $ will stand for the Schwartz space of 
$\, C^{\; \infty}\, $ functions of rapid decrease on~$\, \mathbb {R}^{\, 3\, }$, 
$\, \mathscr {D\; }(\, \mathbb {R}^{\, 3}\, )\, $ for the space of 
$\, C^{\; \infty}\, $ functions of compact support.
Furthermore, we denote by $\, \mathscr {A}_{\; obs}\, $ the observable algebra of the model and use the symbol $\, \mathscr {A}^{\,\, e.\, m.}\, $ for  the 
subalgebra generated by $\, F^{\; \mu\, \nu}\, (\, \mathbf {x}\, ,\, t\, )\, $.\\  
The support of the convolution of $\, \rho\, $ with a given charged-particle wave function is denoted 
by $\, \mathcal {O}\, $ and its causal complement by $\, \mathcal {O}\, '.\, $
The symbol $\, \mathcal {O}_{\, +}^{}\, $ will stand for the region of Minkowski space formed by the points 
which have positive time-like distance from all the points of $\, \mathcal {O}\, ;$ such a region will be 
referred to as the future tangent of~$\, \mathcal {O}\, ,\, $ the past tangent $\, \mathcal{O}_{\; -}^{}
\, $ being defined likewise, with obvious replacements.
The algebra of observables associated to a given region $\, \mathscr {C}\, $ of 
Minkowski space is denoted by $\, \mathscr {A}\, (\, \mathscr {C}\, )\, .$
\newline

The system that we shall consider consists of a single charged non-relativistic quantum particle coupled 
to the quantum electromagnetic field and its dynamics is governed by the Hamiltonian 
\begin {equation}\label {hamiltoniano Feynman}
H^{\,\, (\, v\, )}\, \equiv\,\,\, \hat {\mathbf {p}}\; \cdot\; \mathbf {v}\;\, +\; H_{\,\, 0}^{\,\, e.\, m.}\, +
\, e\;\, v\, \cdot\, A\,\, (\, \rho\, ,\, \hat {\mathbf {x}}\, )\; \equiv\,\, H_{\,\, 0}^{\,\, (\, v\, )\; }+\; 
H_{\; int\; ,\;\, \lambda}^{\,\, (\, v\, )\; ,\;\, (\, \hat {\mathbf {x}}\, )}\;\, ,
\end {equation}
with $\, v\, \equiv\, (\, 1\, ,\, \mathbf {v}\, )\, $ and $\, \rho\, \in\mathscr {S\, }(\, \mathbb {R}^{\, 3}\, )\, $ a 
rotationally invariant distribution of charge, serving as ultraviolet cutoff.\\
The Heisenberg equations governing the dynamics of the algebra of the four-vector potential of 
the $\, BN\, $ model are
\begin {equation}\label {dinamica del potenziale}
\Box\,\, A^{\,\, \mu}\,\, (\, \mathbf {x}\, ,\; t\, )\, =\,\, j_{\,\, \hat {\mathbf {y}}}^{\;\, \mu}
\; (\, \mathbf {x}\, ,\; t\, )\; ,\quad\quad\quad\quad\quad\quad\quad
\end {equation}
where $\, j^{\,\, \mu}\, $ is the conserved charge-current density given by 
\begin {equation}\label {corrente conservata Gupta}
j_{\,\, \hat {\mathbf {y}}}^{\,\, \mu}\; (\, \mathbf {x}\, ,\; t\, )\, =\,\, e\,\,\, (\,\, \theta\; (\, -\, t\, )\,\,\, v^{\; \mu}
\,\,\, \rho\; (\, \vert\, \mathbf {x}\, -\, \hat {\mathbf {y}}\, -\, \mathbf {v}\;\, t\, \vert\, )\,\, +
\;\, \theta\; (\, t\, )\;\,\, v\; '^{\; \mu}\,\,\, \rho\,\, (\, \vert\, \mathbf {x}\, -\, \hat {\mathbf {y}}\, -
\, \mathbf {v}\; '\,\, t\, \vert\, )\, )\; .
\end {equation}
$\mathbf {v}\, $ is a triple of self-adjoint operators in a Hilbert space, to be identified as the observable 
corresponding to the asymptotic velocity of the particle. 
They commute with the Weyl algebra $\, \mathscr {A}_{\; ch}^{}\, $ generated by the canonical variables of 
the electron and with the polynomial algebras generated, in the Coulomb and Feynman gauge respectively, 
by the photon canonical variables.
In the following, it will not be necessary to specify the detailed form of the interaction which 
changes the value of the $\, \mathbf {v}\, $- operators.

By taking the Fourier-transform, we can write down and easily solve (\ref {dinamica del potenziale}),
(\ref {corrente conservata Gupta}) in energy-momentum space. 
The equation of motion for the annihilation operator-valued distribution is
\begin {eqnarray}\label {equazione equivalente}
i\;\; \frac {\, d\,\, a^{\; \mu\, }\, (\, \mathbf {k}\, ,\; t\, )} {d\;\, t}\, =\,\, \vert\, \mathbf {k}\, \vert
\;\,\, a^{\; \mu\, }\, (\, \mathbf {k}\, ,\; t\, )\, -\frac {\; \tilde {j\, }_{\hat {\mathbf {y}}}^{\; \mu\; }
(\, \mathbf {k}\, ,\; t\, )} {\sqrt {\; 2\;\, \vert\, \mathbf {k}\, \vert}\,\, }
\,\,\, ,
\end {eqnarray}
with
\begin {equation}
\tilde {j\; }_{\hat {\mathbf {y}}}^{\; \mu\,\, }(\, \mathbf {k}\, ,\; t\, )\, =\;\, e\,\,\, \tilde {\rho\,\, }(\, \mathbf {k}\, )
\,\, (\,\, \theta\; (\, t\, )\;\,\, v\; '^{\,\, \mu}\,\,\, e^{\; -\; i\,\,\, \mathbf {k}\; \cdot\; (\,\, \hat {\mathbf {y}}\,\, +
\,\, \mathbf {v}\, '\,\, t\; )}\, +\; 
\theta\,\, (\, -\; t\; )\,\,\, v^{\;\, \mu}\;\,\, e^{\; -\; i\,\,\, \mathbf {k}\; 
\cdot\; (\,\, \hat {\mathbf {y}}\,\, +\,\, \mathbf {v}\;\, t\; )}\,\, )\; .
\nonumber
\end {equation}
Since equation (\ref {equazione equivalente}) is linear and non homogeneous, its solution can be written as 
the sum of the general solution of the free equation and of a solution of the non-homogeneous equation. 
For positive times,
\begin {eqnarray}\label {equazione per a}
a^{\; \mu\; }(\, \mathbf {k}\, ,\; t\, )\, =\,\,  e^{\; -\, i\,\, \vert\, \mathbf {k}\, \vert\,\, t\; }\,\,
[\,\, a^{\; \mu\,\, }(\, \mathbf {k}\, ,\; t_{\; 0}^{}=\, 0\, )\, +
\frac {e\;\, v\,\, '^{\,\, \mu}\;\, \tilde {\rho\,\, }(\, \mathbf {k}\, )\, } 
{\sqrt {\,\, 2\; \vert\, \mathbf {k}\, \vert\, }\; }
\;\, \frac {\, e^{\,\, i\;\, v\; '\, \cdot\,\, k\;\,\, t}\, -\, 1\, } 
{v\; '\cdot\; k\, }
\;\, e^{\; -\, i\,\,\, \mathbf {k}\; \cdot\,\, \hat {\mathbf {y}}}\;\, ]
\quad\quad\nonumber\\
\equiv\,\, a_{\,\, 0}^{\,\, \mu}\; (\, \mathbf {k}\, ,\; t\, )\, +\:
f_{\; v\: '}^{\,\, \mu}\, (\, \mathbf {k}\, ,\; t\; ;\, 
\hat {\mathbf {y}}\, )\; .\quad\;\,
\end {eqnarray}
In order to obtain the solution for $\, t<0\, ,\, $ it suffices to replace the final value of the four-velocity operator 
by the initial one.
In the sequel, for definiteness we shall state the results for positive times.
The space-time dependence of the four-vector potential can be computed by means of Fourier transformation, 
employing (\ref {equazione per a}); one has
\begin {equation}\label {soluzione del modello covariante}
A^{\,\, \mu}\,\, (\, \mathbf {x}\, ,\; t\, )\, =\; A_{\,\, 0}^{\,\, \mu}\; (\, \mathbf {x}\, ,\; t\, )\, +
\; e\;\, F_{\,\, v\: '}^{\:\, \mu}\; (\, \mathbf {x}\, ,\; t\; ;\, \hat {\mathbf {y}}\, )\; ,
\quad\quad\quad\quad
\end {equation}
where $\, A_{\,\, 0}^{\,\, \mu}\, $ satisfies $\, \Box\; A_{\,\, 0}^{\,\, \mu}\; (\, \mathbf {x}\, ,\; t\, )\, =\; 0\, $
and the $\, CCR\, $ 
\begin {equation}\label {regole di commutazione canoniche per i potenziali}
[\,\, A_{\,\, 0}^{\,\, \mu\,\, }(\, \mathbf {x}\, ,\; t\, )\; ,\; \dot {A}_{\,\, 0}^{\;\, \nu}\; (\, \mathbf {x}\; ',\; t\, )\,\, ]\, =
\, -\,\, i\,\, g^{\; \mu\; \nu}\;\,\, \delta\; (\, \mathbf {x}\, -\, \mathbf {x}\; '\, )\; ,\quad\quad\quad\quad\quad\quad
\end {equation}
and $\, F^{\,\, \mu}\, $ is the (operator-valued) function
\begin {eqnarray}\label {contributo numerico}
F_{\,\, v\; '}^{\,\, \mu}\, (\, \mathbf {x}\, ,\; t\; ;\, \hat {\mathbf {y}}\, )\, =
\; \frac {1} {(\, 2\; \pi\, )^{\,\, 3\, /\; 2}\, }\;\, \int\;\, d^{\;\, 3\; }k\,\,\,
\,\, e^{\,\, i\,\,\, \mathbf {k}\: \cdot\,\, \mathbf {x}}\;\;\,\, 
\tilde {F}_{\,\, v\; '}^{\,\, \mu}\; (\, \mathbf {k}\, ,
\; t\; ;\, \hat {\mathbf {y}}\, )\; ,
\quad\quad\;\; 
\nonumber\\
\\
\;\; \tilde {F}_{\,\, v\; '}^{\,\, \mu}\, (\, \mathbf {k}\, ,\; t\; ;\, \hat {\mathbf {y}}\, )
\, \equiv\,\, \frac {\, 1} {\sqrt {\,\, 2\,\, \vert\, \mathbf {k}\, \vert}\, }\;\,\, 
[\;\, f_{\,\, v\: '}^{\,\, \mu}\, (\; \mathbf {k}\, ,\; t\; ;\, \hat {\mathbf {y}}\, )\, +
\; \overline {f\, }_{v\; '}^{\;\, \mu}\,\, (\, -\, \mathbf {k}\, ,\; t\; ;\, 
\hat {\mathbf {y}}\, )\; ]\; ,\;\;\; 
\nonumber
\end {eqnarray}
where $\, f_{\,\, v\: '}^{\,\, \mu}\, $ has been defined in (\ref {equazione per a}).\\
The expression of $\, F_{\,\, v\; '}^{\,\, \mu}\, $ for $\, \textbf {v}\; '=\, 0\,\, $ is denoted by $\,  F_{\;\, v_{\; 0}^{}}^{\,\, \mu}\, $ and is calculated below.
One has  
\begin {equation}\label {contributo numerico velocita' nulla}
F_{\,\, v_{\; 0}^{}}^{\,\, \mu}\, (\, \mathbf {x}\, ,\; t\; ;\; \hat {\mathbf {y}}\, )\, =
\; \frac {1} {(\, 2\; \pi\, )^{\,\, 3\, /\; 2}\, }\;\, \int\;\, d^{\,\, 3\, }k\;\;\, 
e^{\;\, i\,\,\, \mathbf {k}\; \cdot\; (\, \mathbf {x}\, -
\; \hat {\mathbf {y}}\, )}\;\,\, \tilde {\rho\,\, }
(\, \mathbf {k}\, )\,\,\, 
\frac {\, 1\, -\; \cos\, \vert\, \mathbf {k}\, \vert\;\, t\, } 
{{\vert\, \mathbf {k}\, \vert}^{\,\, 2}\, }\,\,\,\,\, 
\delta^{\;\, \mu\; 0}\;\, .
\end {equation}
In order to determine the localization of the support of (\ref{contributo numerico velocita' nulla}),
we note that the definition of the characteristic function $\, \chi_{\,\, I}^{}\, $ of the interval $\, I\, $ of the real line,
\begin {equation}
\chi_{\;\, [\,\, -\,\, x\; ,\;\, x\;\, ]}^{}\,\, (\; y\; )\; =\,\, \frac {1} {\; \pi\;\, }\;\, \int_{\; -\; \infty}^{\;\, +\; \infty}
\;\, d\;\, \xi\;\;\,\, e^{\; -\, i\,\,\, \xi\,\,\, y}\;\;\;\, \frac {\, \sin\,\, \xi\,\, x\, } {\xi\, }\;\, ,\;\, x\, >\, 0\,\, ,
\quad
\end {equation}
implies the equality
\begin {equation}\label {definizione funzione caratteristica}
\frac {1} {(\, 2\; \pi\, )^{\;\, 3\, /\; 2}\, }\,\, \int\,\, d^{\,\, 3\; }k\;\;\, e^{\,\, i\,\,\, \mathbf {k}\; \cdot\,\, \mathbf {x}}
\;\,\,\,\, \frac {\, 1\, -\; \cos\, \vert\, \mathbf {k}\, \vert\;\, t\, } {{\vert\: \mathbf {k}\, \vert}^{\,\, 2}\, }\; =
\;\, \sqrt {\; \frac {\pi} {2}}\;\,\,\, \frac {\, \chi_{\,\, \vert\, \mathbf {x}\, \vert\; <\,\, \vert\, t\, \vert}^{}} 
{\vert\, \mathbf {x}\, \vert}\;\, \cdot\quad
\end {equation}
Therefore (\ref {contributo numerico velocita' nulla}) can be cast in the form
\begin {equation}\label {F a riposo}
F_{\,\, v_{\; 0}^{}}^{\,\, \mu}\; (\, \mathbf {x}\, ,\; t\; ;\, \hat {\mathbf {y}}\, )\, =\,\, \int\,\, d^{\,\,\, 3\, }z
\;\,\,\, \rho\,\, (\, \vert\, \mathbf {x}\, -\, \hat {\mathbf {y}}\, -\, \mathbf {z}\, \vert\, )\;\,\, 
\chi_{\,\, \vert\, \mathbf {z}\, \vert\; <\,\, \vert\, t\, \vert}^{}\;\,\, \frac {\, \delta^{\;\, \mu\,\, 0\, }} 
{4\,\, \pi\;\, \vert\, \mathbf {z}\, \vert\, }\;\, \cdot\quad
\end {equation}
The result for a non-vanishing value $\, \textbf {v}\, '\, $ of the $\, \textbf {v}\; $- operator could be obtained at once by the Lorentz covariance 
of the Gupta-Bleuler formulation, were it not for the fact that a non-covariant ultraviolet cutoff has been 
employed in the model.
By introducing a Dirac delta, one can write the right-hand side (r.h.s.) of (\ref {F a riposo}) as an integral over 
a volume element in Minkowski space; a change of integration variables in the resulting expression, 
performed by means of a boost corresponding to the four-velocity $\, v\,\, ',\, $ then yields
\begin {align}\label {funzione F in soluzione Gupta}
F_{\,\, v\; '}^{\,\, \mu}\, (\, \mathbf {x}\, ,\; t\; ;\; \hat {\mathbf {y}}\, )\, =\; \int\,\, d^{\;\, 4\, }z\,\,\,\, 
\rho\,\, (\, \vert\, \mathbf {x}\, -\, \hat {\mathbf {y}}\, -\, \underline {\Lambda_{\; v\; '}^{}
\,\, z}\, \vert\, )\;\,\, \delta\,\, (\, (\, \Lambda_{\; v\; '}^{}\,\, z\, )_{\,\, 0}^{}\, )\;\,
\quad\quad\quad\nonumber\\
\times\,\,\, \frac {{v\,\, '}^{\,\, \mu}\,\,\,\, \chi_{\,\, z^{\, 2}\,\, >\;\, 0}^{}} 
{4\,\, \pi\,\,\, [\; (\,\, v\,\, '\, \cdot\, z\; )^{\,\, 2}\, -\, 
{v\,\, '}^{\,\, 2}\,\, z^{\; 2}\;\, ]^{\,\, 1\, /\; 2}\, }\,\, \cdot
\end {align}
Apart from the convolution with the $\, \rho\, $ function and from the presence of the Dirac delta, which accounts for 
the non covariance of the high-energy cutoff, the above expression equals the Li\'enard-Wiechert potential 
generated by a charge moving with a constant velocity $\, \mathbf {v}\: '\, ;\, $ for negative times, the retarded potential is 
replaced by the advanced one.

We can now define and evaluate the asymptotic e.m. fields.
Given a solution $\, g\, $ of the free wave equation, such that 
$\, g\; (\, \mathbf {x}\, ,\, t\, )\in\mathscr {S}\, (\, 
{\mathbb {R}}^{\; 3}\, )\, $ $\, \forall\,\, t\; ,\, $ we 
define 
\begin {equation}
A^{\,\, \mu}\, (\, g\, )\; (\, x_{\; 0}^{}\, ,\; t\, )\, \equiv\; \int\,\, d^{\,\, 3\, }x\;\,\,\, 
\overline {g\, }\; (\, \mathbf {x}\, ,\; x_{\; 0}^{}\, )\,\,\, \overleftrightarrow 
{\;\, \partial_{\;\, x_{\; 0}^{}}}\;\,\, A^{\,\, \mu}\; (\, \mathbf {x}\, ,\; x_{\,\, 0\, }^{}+
\; t\, )
\quad\quad
\end {equation} 
and the corresponding smearing of the (operator-valued) function, which appears in the solution 
(\ref {soluzione del modello covariante}),
\begin {equation}\label {media Klein Gordon}
F_{\,\, v}^{\,\, \mu}\, (\, g\, )\; (\, x_{\,\, 0}^{}\, ,\; t\; ;\, \hat {\mathbf {y}}\, )\, \equiv
\, \int\,\, d^{\;\, 3\; }x\;\,\,\, \overline {g\, }\; (\, \mathbf {x}\, ,\; x_{\,\, 0}^{}\, )
\; \overleftrightarrow {\;\, \partial_{\;\, x_{\; 0}^{}}}\;\,\, F_{\,\, v}^{\,\, \mu}\;
(\, \mathbf {x}\, ,\; x_{\,\, 0\, }^{}+\; t\; ;\, \hat {\mathbf {y}}\, )\; ,
\end {equation} 
where $\, f\: \overleftrightarrow {\; \partial_{\,\, x_{\: 0}^{}}}\,\, g\equiv\, f\,\,
\partial_{\,\, x_{\: 0}^{}}\: g\, -\, (\, \partial_{\,\, x_{\: 0}^{}}\: f\, )\,\, g\, .\, $
We give the calculation for $\, \textbf {v}=0\, $ and for the $\, out\, $- field, and write for brevity $\, F_{\,\, v_{\; 0}^{}}^{}\equiv F_{\,\, v_{\: 0}^{}}^{\,\, \mu\: =\,\, 0\,\, }.\, $
Equations (\ref {F a riposo}), (\ref {media Klein Gordon}), the properties of the test function and the 
Riemann-Lebesgue lemma lead to the existence of the limit (in the strong topology of multiplication 
operators)
\begin {eqnarray}\label {calcolo campo asintotico}
F_{\;\, v_{\; 0}^{}}^{\,\, out}\, (\, g\, )\; (\, t\; ;\, \hat {\mathbf {y}}\, )\, \equiv\; \lim_{x_{\; 0}^{}\,\, \rightarrow\; +\; \infty}
\; F_{\,\, v_{\; 0}^{}}^{}\, (\, g\, )\; (\, x_{\,\, 0}^{}\, ,\; t\; ;\, \hat {\mathbf {y}}\, )\, =\, -\int\,\, d^{\,\, 3\, }k\,\,\,\, 
e^{\; -\, i\;\, \mathbf {k}\; \cdot\; \hat {\mathbf {y}}}
\quad\quad\nonumber\\
\times\;\, \tilde {\rho\; }(\, \mathbf {k}\, )\,\,\, [\,\,\, \overline {\tilde {g\; }}\,
(\, \mathbf {k}\, ,\; x_{\; 0}^{}\, )\;\,\, \overleftrightarrow 
{\,\,\, \partial_{\;\, x_{\: 0}^{}}}\;\;\,\, \frac {\, \cos\,
\vert\, \mathbf {k}\, \vert\,\, (\, x_{\; 0\, }^{}+\; t\, )} 
{{\vert\, \mathbf {k}\, \vert}^{\,\, 2}}\,\,\, ]\, 
\rvert_{\;\, x_{\; 0}^{}\,\, =\,\,\, 0}^{}\;
\quad\quad\nonumber\\
\equiv\int\,\, d^{\,\, 3\, }x\,\,\, [\,\,\, \overline {g\, }\; (\, \mathbf {x}\, ,
\; x_{\,\, 0}^{}\, )\,\,\, \overleftrightarrow {\;\, \partial_{\;\, x_{\: 0}^{}}}\;\;
\, G_{\,\, v_{\: 0}^{}\; }^{}(\, \mathbf {x}\, ,\; x_{\,\, 0\, }^{}+\; t\; ;\,
\hat {\mathbf {y}}\, )\; ]\, \rvert_{\;\, x_{\; 0}^{}\,\, =\,\,\, 0}^{}\;\, ,
\end {eqnarray}
with
\begin {equation}\label {campo asintotico}
G_{\,\, v_{\: 0}^{}}^{}\, (\, \mathbf {x}\, ,\; t\, ;\, \hat {\mathbf {y}}\, )\, =\, -
\int\,\, d^{\,\, 3\, }z\,\,\,\, \rho\,\, (\, \vert\, \mathbf {x}\, -\, 
\hat {\mathbf {y}}\, -\, \mathbf {z}\, \vert\, )\;\;\, 
\frac {\; \chi_{\,\, \vert\, \mathbf {z}\, \vert\,\, >\;\, 
\vert\, t\, \vert}^{}\; } {4\,\, \pi\,\, \vert\,
\mathbf {z}\, \vert}\; \cdot
\quad\quad\quad\quad\quad
\end {equation}
It thus follows the existence of the $\, out\; $- field
\begin {eqnarray}
A_{\,\, out}^{\,\, \mu}\, (\, g\, )\; (\, t\, )\, \equiv\; \lim_{x_{\; 0}^{}\,\, \rightarrow\; +\; \infty}\; 
A^{\,\, \mu}\; (\, g\, )\; (\, x_{\,\, 0}^{}\, ,\; t\, )\,\, \equiv\,\, \int\,\, d^{\,\, 3\, }x\;\,\, 
[\;\,\, \overline {g}\,\, (\, \mathbf {x}\, ,\; x_{\,\, 0}^{}\, )\quad\quad
\nonumber\\
\times\;\, \overleftrightarrow {\;\, \partial_{\;\, x_{\; 0}^{}}}\;\;\, A_{\,\, out}^{\,\, \mu}\; (\, \mathbf {x}\, ,
\; x_{\,\, 0\, }^{}+\; t\; )\; ]\, \rvert_{\;\, x_{\,\, 0}^{}\,\, =\,\,\, 0}^{}\;\, ,
\end {eqnarray}
\begin {equation}\label {campo asintotico in rappresentazione Gupta}
A_{\,\, out\,\,\, }^{\,\, \mu}(\, \mathbf {x}\, ,\; t\, )\, =\,\, A_{\; free\,\, }^{\,\, \mu}(\, \mathbf {x}\, ,\; t\, )\, +\; 
G_{\,\, v_{\: 0}^{}}^{\,\, \mu}\, (\, \mathbf {x}\, ,\; t\; ;\, \hat {\mathbf {y}}\, )\; ,
\quad\quad\quad\quad\quad\quad\quad\quad\quad\quad\quad
\end{equation}
with $\, G_{\,\, v_{\: 0}^{}}^{\,\, \mu}\equiv\, G_{\; v_{\; 0}^{}\; }^{}\; \delta^{\;\, \mu\; 0}\,\, .$ 

We shall denote by $\, \mathscr {F}^{\,\, as}\, $ the algebra of the electromagnetic observables constructed 
in terms of the asymptotic vector-potential $\, A_{\,\, as\; }^{},\, as\, =\, in,out\, $.

\section {Space-Time Properties of The Local Formulation}

The aim of this Section is to give an algebraic formulation to the relativistically covariant dynamics
given by eqs.(\ref {dinamica del potenziale}), (\ref {corrente conservata Gupta}) and to analyze
the space-time properties of the expectations of product functionals, corresponding to vectors
of the Gupta-Bleuler space of the model.

We recall that in the $\, FGB\, $ gauge an observable is defined as a local function of the gauge fields which is left 
pointwise invariant by the residual symmetry group of the theory, the so-called gauge transformations of 
the second kind\cite {Sym71}. 
Equivalently, it can be identified by the condition that it commutes with the generator of such transformations, 
the $\, \partial\, \cdot A\, $ field, which is an observable in the above sense.
In the sequel, it will be also denoted by $\, B\, .\, $\\
In the subsequent analysis it will be useful to consider the free electromagnetic algebras
$\, \mathcal {F}_{\,\, 0}^{}\, $ and $\, \mathscr {A}_{\;\, 0}^{\,\, e.\, m.}\, $ generated respectively by $\, A_{\,\, 0}^{\,\, \nu}\, (\, \mathbf {x}\, ,\, t\, )\, $ and 
$\, F_{\,\, 0}^{\; \mu\; \nu\; }(\, \mathbf {x}\, ,\; t\, )\equiv\, (\; \partial^{\,\, \mu\, }
\, A_{\,\, 0}^{\;\, \nu\, }-\, \partial^{\,\, \nu\, }\, A_{\,\, 0}^{\,\, \mu}\; )\,\, 
(\, \mathbf {x}\, ,\; t\, )\, ,\, $
which is left invariant by the gauge transformations of the second kind of the non-interacting theory and 
thus commutes with their generator $\, B_{\,\, 0}^{}\equiv\, \partial\, \cdot\, A_{\,\, 0}^{}\, .\, $\\ 
The elements of $\, \mathscr {A}_{\;\, 0}^{\;\, e.\, m.}\, $ thus fulfill 
\begin {align}
\partial_{\,\, \mu\, }^{}\; F_{\,\, 0}^{\,\, \mu\; \nu}\, (\, \mathbf {x}\, ,\; t\, )\, =
\, -\; \partial^{\,\, \nu}\,\, B_{\,\, 0}^{}\; (\, \mathbf {x}\, ,\; t\, )\; ,\;\,
\Box\,\, B_{\,\, 0}^{}\; (\, \mathbf {x}\, ,\; t\, )\, =\,\, 0\; ,
\quad\nonumber\\
\epsilon_{\; \mu\,\, \nu\,\, \rho\,\, \sigma}^{}\,\,\, \partial^{\,\, \nu\, }\,\, F_{\,\, 0}^{\,\, \rho\,\, \sigma}
\, (\, \mathbf {x}\, ,\; t\, )\, =\,\, 0\,\, ,\,\, [\,\, F_{\,\, 0}^{\,\, \rho\; \sigma}\, (\, x\, )\; ,\, 
B_{\,\, 0}^{}\; (\, x\; '\, )\,\, ]\; =\,\, 0\,\, ,
\end {align}
and the equal time canonical commutation relations induced by 
(\ref {regole di commutazione canoniche per i potenziali}).
Moreover, owing to (\ref {soluzione del modello covariante}), the generators of the restricted gauge 
transformations for $\, \mathscr {A}^{\;\, e.\, m.}\, $ and $\, \mathscr {A}_{\,\,\, 0}^{\;\, e.\, m.}\, $ are 
related by
\begin {equation}\label {relazione tra i generatori}
B\; (\, \mathbf {x}\, ,\; t\, )\, =\,\, B_{\,\, 0}^{}\, (\, \mathbf {x}\, ,\; t\, )\, +\; e\,\,
(\, \partial\, \cdot\, F\, )\: (\, \mathbf {x}\, ,\; t\,\, ;\; \hat {\mathbf {y}}\, )\; .
\quad\quad\,\,
\end {equation} 
For $\, t\, >\, 0\; $ we also get the relation
\begin {align}\label {osservabili nel modello}
F^{\,\, \mu\; \nu}\; (\, \mathbf {x}\, ,\; t\, )\, =\,\, F_{\;\, 0}^{\,\, \mu\; \nu\, }\, (\, \mathbf {x}\, ,\; t\, )\, +
\, H_{\; v\; '}^{\,\, \mu\; \nu\, }\, (\, \mathbf {x}\, ,\; t\; ;\, \hat {\mathbf {y}}\, )\; ,
\quad\quad\quad\nonumber\\
\\
H_{\; v\; '}^{\,\, \mu\; \nu\, }\, (\, \mathbf {x}\, ,\; t\; ;\, \hat {\mathbf {y}}\, )\, \equiv\; (\; \partial^{\,\, \mu}
\; F_{\,\, v\; '}^{\,\, \nu\, }-\, \partial^{\,\, \nu}\; F_{\,\, v\; '}^{\,\, \mu}\, )\,\, (\, \mathbf {x}\, ,\; t\; ;\, 
\hat {\mathbf {y}}\, )\; .
\quad\;\; \nonumber
\end {align}

It is useful to recall that within the perturbative-theoretic treatment of $\, QED\, $ in a local and covariant 
gauge the vacuum representation is required to be positive on the observables, as in a quantum 
field theory with positive-definite Wightman correlation functions.
The outcome is an expansion around a non-interacting theory (in terms of renormalized parameters) and the 
vacuum representation is characterized by the choice of the Fock representation for the free four-vector 
potential.

Accordingly, in the model we shall assume a Fock representation for $\, A_{\,\, 0}^{\;\, \mu}\, .\, $
For a given single particle state $\, \omega_{\; ch}^{}\, ,\, $ we consider the product functional 
\begin {equation}\label {funzionale di Gupta}
\omega_{\,\, G}^{}\, \equiv\,\, \omega_{\,\, ch}^{}\, \otimes\; \omega_{\,\, F}^{}\; ,\,\,
\quad\quad\quad\quad\quad
\end {equation}
acting on the algebras $\, \mathscr {A}_{\,\, ch}^{}\, $ and $\, \mathcal {F}_{\,\, 0}^{}\, $.\\ 
The space obtained via the $\, GNS\, $ construction on $\, \omega_{\,\, G}^{}\, $ is 
\begin {equation}
\mathscr {G}\, \equiv\,\, L^{\,\, 2\,\, }(\, \mathbf {y}\, ;\, v\, )\, \otimes\; \mathscr {D}\;\, ,
\quad\quad\quad\quad
\end {equation}
with $\, L^{\;\, 2}\, $ the one-particle Hilbert space and $\, \mathscr {D}\, $ the indefinite-metric Fock space 
obtained by applying polynomials of the free four-vector potential to the Fock vacuum.\\
Positivity of $\, \omega_{\,\, F}^{}\,$ on $\, \mathscr {A}_{\;\, 0}^{\;\, e.\, m.}\, $ implies that any vector belonging to the subspace of $\, \mathscr {G}\, $ given by
\begin {equation}\label {sottospazio positivo di Gupta nel modello}
\mathcal {K}\,\, \equiv\;\, L^{\,\, 2\;\, }(\, \mathbf {y}\, ;\, v\, )\; \otimes
\,\, \mathscr {A}_{\;\, 0}^{\;\, e.\, m.}\,\,\, \Psi_{\: F}^{}\quad\quad
\end {equation}
also defines a positive functional on $\, \mathscr {A}_{\;\, 0}^{\;\, e.\, m.}\, $.

Moreover, $\, \mathcal {K}\, $ is a space of physical states for $\, \mathscr {A}_{\;\, 0}^{\;\, e.\, m.}\, ,\, $ 
since
\begin {equation}
\partial^{\;\, \mu\, }\,\, F_{\,\, 0\,\, ,\;\, \mu\; \nu\, }^{\,\,\, (\, -\, )}\, (\, \mathbf {x}\, ,\; t\, )\,\,\, \Psi_{\: F\, }^{}=\,\, 0
\quad\quad\quad\quad\;
\end {equation}
implies
\begin {equation}\label {condizione Gupta libera}
\partial^{\;\, \mu\, }\,\, F_{\,\, 0\,\, ,\;\, \mu\; \nu\, }^{\,\,\, (\, -\, )}\, (\, \mathbf {x}\, ,\; t\, )
\,\,\, \phi\; =\,\, 0\,\, ,\,\, \forall\,\, \phi\, \in\, \mathcal {K}\,\, ,
\end {equation}
and by the positivity of $\, \mathcal {K}\, $ one also has
\begin {equation}\label {relazione operatoriale Gupta-Bleuler} 
\partial^{\;\, \mu\, }\,\, F_{\,\, 0\; ,\,\, \mu\; \nu\, }^{}\; (\, \mathbf {x}\, ,\; t\, )\, =\,\, 0
\quad\quad\quad\quad\quad\quad\quad
\end {equation}
in $\, \mathcal {K}\, .\, $

It is important to stress that the functionals $\, \omega_{\,\, G}^{}\, $ are positive on $\, \mathscr {A}_{\; obs}^{}\, ,\, $ 
due to the fact that they are product functionals and to the explicit expression of the electromagnetic 
observables, given by (\ref {osservabili nel modello}).
Of course, $\, \mathcal {K}\, $ is not a space of physical states for $\, \mathscr {A}\, ,\, $ because the time 
evolution of the observable algebra does not obey the Maxwell equations; the deviation from Gauss' law 
is in fact given by
\begin {equation}\label {deviazione Gupta da Gauss}
\omega_{\,\, G\, }^{}\, (\,\, \partial_{\,\, \mu}^{}\,\, F^{\,\, \mu\; \nu}\; (\, x\, )\, )\, =
\;\, \omega_{\; ch}^{}\; (\, j_{\; \hat {\mathbf {y}}}^{\,\, \nu}\; (\, x\, )\, -
\; e\,\,\, \partial^{\,\, \nu}\; (\, \partial^{}\, \cdot\, F\, )\; (\, x\: ;\:
\hat {\mathbf {y}}\, )\, )\; .\;\;\;
\end {equation}
The previous relation holds for an arbitrary charged particle state, due to the fact that the negative-frequency 
component of the representative of $\, B\, $ acts as a multiplication on the $\, L^{\,\, 2}\, $ space of states of the charge; as a 
matter of fact, it follows from (\ref {sottospazio positivo di Gupta nel modello}), 
(\ref {relazione tra i generatori}) that $\, \forall\,\, \psi\, \in\, L^{\,\, 2}\, $
\begin {equation}\label {autovalore condizione ausiliaria}
B^{\,\, (\, -\, )}\; (\, \mathbf {x}\, ,\; t\, )\,\, (\, \psi\, (\, \mathbf {y\, })\, \otimes\, \mathscr {A}_{\:\, 0}^{\;\, e.\, m.}
\;\, \Psi_{\: F}^{}\, )\, =\,\, e\;\, (\, \partial\, \cdot\, F\, )^{\,\, (\, -\, )}\,\, (\, \mathbf {x}\, ,\; t\, ;\, \mathbf {y}\, )
\,\, (\, \psi\, (\, \mathbf {y\, })\, \otimes\, \mathscr {A}_{\:\, 0}^{\;\, e.\, m.}\,\, \Psi_{\: F}^{}\, )\: .
\end {equation}
The space-time localization of the support of the eigenvalue on the r.h.s. of eq.(\ref {autovalore condizione 
ausiliaria}) will be relevant in the characterization of the physical charged states carried out in the next 
Section; hence it is convenient to evaluate its explicit expression.
It follows from (\ref {contributo numerico velocita' nulla}) that
\begin {equation}\label {supporto autovalore Gupta}
(\, \partial\, \cdot\, F\, )^{\,\, (\, -\, )}\,\, (\, \mathbf {x}\, ,\; t\; ;\, \mathbf {y}\, )\, =
\,\, \int\,\, d^{\,\,\, 3\, }z\;\;\, \rho\,\, (\, \vert\, \mathbf {x}\, -\, \mathbf {y}\, -\,
\mathbf {z}\, \vert\; )\;\;\, \Delta^{\; (\, -\, )}\,\, (\, \mathbf {z}\, ,\; t\, )\; ,\;
\quad
\end {equation}
where $\, \Delta^{\; (\, -\, )}\; (\, x\, )\, $ is related to the Pauli-Jordan distribution 
 \begin {equation}
\Delta\,\, (\, \mathbf {x}\, , \; t\, )\, =\,\, \frac {1} {2\,\, \pi\, }\;\;\, \epsilon\; (\, t\,  ) 
\;\,\, \delta\; (\; t^{\; 2\, }-\, \mathbf {x}^{\; 2}\; )
\quad\quad\quad
\end {equation}
by $\, \Delta\,\, (\, x\, )\, =\: \Delta^{\; (\, -\, )}\,\, (\, x\, )-\, \Delta^{\,\, (\, -\, )}\,\, (\, -\, x\, )\: $.

Equations (\ref {deviazione Gupta da Gauss}), (\ref {supporto autovalore Gupta}) imply that Gauss' law 
is fulfilled by the restriction of the functionals (\ref {funzionale di Gupta}) to the observables associated 
to $\, \mathcal{O}_{\, +\, }^{};\, $ it follows that $\, \omega_{\,\, G}^{}\, $ acts as a physical state on $\, \mathscr {A}\, 
(\mathcal {\, O}_{\, +\, }^{})\, $, since it is positive on $\, \mathscr {A}_{\,\, obs}^{}\, .\, $

We remark that the validity of the weak Gauss law for the restriction of Gupta-Bleuler functionals to a 
forward lightcone is a non-perturbative feature of $\, QED\, $ in the $\, FGB\, $ gauge\cite {MoSt1}; the 
positivity of such a restriction remains instead an independent issue.

Concerning the space-like asymptotics in the charged (single-particle) sectors of the model, eqs.(\ref {soluzione 
del modello covariante}) and (\ref {funzione F in soluzione Gupta}) imply, for a functional of the form 
(\ref {funzionale di Gupta}), $\, \omega_{\,\, G\,\, }^{}(\, A^{\; \mu\; }
(\, \mathcal {O}\, '\, )\, )= 0\, .\, $
By the space-time localization properties of the electromagnetic observables, given by 
eqs.(\ref {osservabili nel modello}), (\ref {funzione F in soluzione Gupta}), we get
\begin {equation}
\omega_{\,\, G\,\, }^{}(\; \mathscr {A}^{\,\, e.\, m.}\; (\, \mathcal {O}\; '\, )\, )\, =\,\, 
\omega_{\,\, F\,\, }^{}(\; \mathscr {A}^{\,\, e.\, m.}\; (\, \mathcal {O}\; '\, )\, )\; . 
\quad\quad
\end {equation}
We now turn to evaluate the expectations of the asymptotic radiation fields. 
One has
\begin {eqnarray}
\omega_{\,\, G}^{}\; (\, A_{\,\, out\,\, }^{\,\, \mu}(\, \mathbf {x}\, ,\; t\, )\, )\, =\,\, 
\omega_{\,\, G}^{}\; (\, G_{\; v_{\: 0}^{} }^{\,\, \mu}\, (\, \mathbf {x}\, ,\; t\; ;\, 
\hat {\mathbf {y}}\, )\, )\: .\quad\quad
\end{eqnarray}
For $\, \mathbf {v}=0\, ,\, $ a Gupta-Bleuler product state thus yields expectations of $\, A_{\,\, out}^{\,\, \mu}\, $ in $\, \mathcal {O}\: '\, $ 
which are equal in modulus to the Coulomb potential and have the opposite sign.
The $\, 1\, /\, r\, $ behaviour of (\ref {campo asintotico}) at space-like infinity is related to the fact that states with 
non-zero charge belonging to $\, \mathscr {G}\, $ induce a non-Fock representation of the asymptotic vector potential;
therefore, one can infer the occurrence of infinite photons at asymptotic times even for charged 
non-physical states. 
This result agrees with those obtained on the basis of general hypotheses 
on the structure of local formulations of $\, QED\, $ in Ref.~\onlinecite {FrMoSt79a} as well as in 
Ref.~\onlinecite {MoSt83MoSt84}.

Furthermore, $\, \omega_{\,\, G\,\, }^{}(\, A_{\; out\; }^{\,\, \mu}(\, \mathbf {x}\, ,\, t\, )\, )\, $ vanishes in $\, \mathcal {O}_{\; +}^{}\, \cup\mathcal {\; O}_{\; -}^{}\, $ since, as follows from equation (\ref {F a riposo}), 
there exists a frame of reference in which the interacting field is static in such a region.
Therefore, one has in particular the additional piece of information that the corresponding representation 
of the subalgebra of the asymptotic photon fields constructed with the aid of the outgoing four-vector 
potential in $\, \mathcal {O}_{\; +}^{}\, $ is Fock:
\begin {equation}\label {relazione Fock per asintotici Gupta}
\omega_{\,\, G}^{}\; (\, \mathscr {F}^{\,\, out}\, (\, \mathcal {O}_{\; +}^{}\, )\, )\, =\,\, 
\omega_{\,\, F}^{}\; (\, \mathscr {F}^{\,\, out}\, (\, \mathcal {O}_{\; +}^{}\, )\, )\; .\;\, 
\end {equation}
The consequences of this result for the classification of physical charged 
states will be discussed in Section 3.\\
For a non-vanishing $\, \textbf {v}\, \equiv\; \textbf {v}_{\; out}^{}\; $,
\begin {equation}\label {campo asintotico in rappresentazione Gupta v generica}
A_{\,\, out\,\, }^{\,\, \mu}(\, \mathbf {x}\, ,\; t\, )\, =\,\, A_{\; free\; }^{\,\, \mu}(\, \mathbf {x}\, ,\; t\, )\, +
\, G_{\,\, v_{\; out}^{}}^{\,\, \mu}\, (\, \mathbf {x}\, ,\; t\, ;\, \hat {\mathbf {y}}\, )\; ,
\quad\quad\quad\quad\quad\quad\quad\quad\quad\quad\quad\quad\quad
\end {equation}
\begin {eqnarray}\label {campo asintotico v generica}
\;\; G_{\,\, v_{\, out}^{}}^{\,\, \mu}\, (\, \mathbf {x}\, ,\; t\; ;\, \hat {\mathbf {y}}\, )\, =\, -
\, \int\,\, d^{\;\, 4\; }z\;\,\,\, \rho\,\, (\, \vert\, \mathbf {x\, }-\, \hat {\mathbf {y}}\, -\, 
\underline {\Lambda_{\,\, v_{\, out}^{}}^{}\,\, z}\, \vert\, )\;\,\,\, 
\delta\,\, (\, (\, \Lambda_{\,\, v_{\: out}^{}}^{}\,\, z\, )_{\;\, 0}^{}\,\, )
\quad\quad\quad\quad\quad\nonumber\\
\times\;\, 
\frac {v_{\; out}^{\,\, \mu}\,\,\,\, \chi_{\,\, z^{\: 2}\,\, <\,\,\, 0}^{}} 
{4\,\, \pi\,\,\, [\; (\; v_{\; out}^{}\, \cdot\, z\; )^{\;\, 2}\, -\, 
v_{\; out}^{\,\, 2}\,\,\, z^{\,\, 2}\;\, ]^{\;\, 1\; /\; 2}\, }\; 
\cdot\quad\quad
\end{eqnarray}

Besides (\ref {relazione Fock per asintotici Gupta}), the localization of the support of $\, \omega_{\,\, G\,\, }^{}(\, A_{\; out\; }^{\,\, \mu}(\, \mathbf {x}\, ,\, t\, )\, )\, $ also 
implies the Fock property for the representation of $\, \mathscr {F}^{\; out}\, (\, \mathcal {O}_{\; -}^{}\, )\, ,\, $ confirming results established by Buchholz
\cite {Buch77} for models of field theories with massless bosons and ``standard'' charges (not satisfying 
a Gauss-law type constraint). 
We recall in fact that for such models the representations of boson asymptotic algebras, associated to any 
region of Minkowski-space, either bounded or unbounded, admitting a non-trivial future tangent, have 
been proven to be Fock; the same property is also fulfilled in the $\, BN\, $ model by representations 
of the photon asymptotic algebras associated to the same regions and induced by (product) 
functionals corresponding to vectors of $\, \mathscr {G}\, ,\, $ for which locality, but not 
positivity, holds.
The Fock property of $\, \mathscr {F}^{\; out\,\, }(\, \mathcal {O}_{\; +}^{}\, )\, $ is instead an independent result of the model.
Analogous statements apply to $\, \mathscr {F}^{\,\, in}\, ,\, $ with obvious changes.

For completeness we outline an argument explaining why one expects Fock representations for the asymptotic 
field algebras relative to massless bosons and associated to appropriately chosen regions of Minkowski space. 
Let $\; \mathcal {C}\; $ be a space-time region with a non-trivial future tangent, $\, \pi\, $ an irreducible 
positive-energy representation of the observable algebra and $\, \mathscr {H}\, $ the associated 
$\, GNS\, $ space.
By the Reeh-Schlieder theorem the vacuum is cyclic for the algebra $\; \mathscr {A}\, 
(\, \mathcal {C}_{\; +\, }^{})\, ,\, $ hence the vectors $\, \Psi\, =\, F\; \Omega\, ,$ 
with $\, F \in\mathscr {A}\, (\, \mathcal {C}_{\; +}^{}\, )\, $ and $\, \Omega\, $ 
the ground-state vector, form a dense set $\, \mathscr {V}\, $ of the Hilbert 
space $\, \mathscr {H\, }.$\\ 
On $\, \mathscr {V}\, $ one has $\, A\; \Psi=A\; F\; \Omega=F\; A\; \Omega\, $, with $\, A\, $ belonging to 
the center of $\, \pi\, (\, \mathscr {F}^{\; out}\, (\, \mathcal {C}\, )\, )\, ,$ since by Huyghens' principle 
and locality $\mathscr {F}^{\; out\; }(\, \mathcal {C}\, )\, $ is contained in $\, \mathscr {A}\,'\: 
(\, \mathcal {C}_{\; +\, }^{})\, .\, $
The value taken by the elements of the center of $\, \pi\, (\, \mathscr {F}^{\; out\; }(\, \mathcal {C}\, )\, )\, $ in a factorial representation thus equal those 
of the vacuum representation of the same subalgebra and $\, \pi\, (\mathscr {F}^{\; out}\, (\, \mathcal {C}\, )\, )\, $ is quasi-equivalent to the Fock 
representation.\cite {BuchBos1}

\section {Construction And Properties Of Li\'enard-Wiechert Physical 
Charged States}

As a guide to the construction and of the classification of the charged physical states in the four-vector $\, BN\, $ 
model, first we discuss the same problems in classical electrodynamics.

The electromagnetic fields generated by a charge-current distribution obey Maxwell's equations 
\begin {equation}\label {equazioni di maxwell}
\partial_{\,\, \mu}^{}\;\, F^{\,\, \mu\; \nu}\; =\,\, j^{\;\, \nu}\,\, ,\;
\, \partial^{\;\, \mu}\;\; F_{\,\, \mu\; \nu}^{\;\, *}\, =\,\, 0\;\, ,
\end {equation}
which can be expressed in terms of the vector potential:
\begin {eqnarray}\label {equazioni di Maxwell per i potenziali}
F^{\;\, \mu\,\, \nu}\, =\;\, {\partial}^{\;\, \mu}\,\, A^{\,\, \nu\, }\, -\, {\partial}^{\;\, \nu}\,\, A^{\,\, \mu\, }
\,\, ,\;\,
\nonumber\\
\\
\Box\,\, A^{\,\, \mu\, }\, -\; {\partial}^{\;\, \mu}\;\,\, {\partial}_{\;\, \nu}^{}
\,\,\, A^{\;\, \nu}\, =\;\, j^{\;\, \mu}\;\, .\;\,
\nonumber
\end {eqnarray}
Because of their invariance under gauge transformations, 
\begin {equation}\label {trasformazioni di gauge per i potenziali}
A^{\,\, \mu\, }\, (\, \mathbf {x}\, ,\; t\, )\; \rightarrow\,\, A^{\,\, \mu\, }\, (\, \mathbf {x}\, ,\; t\, )
\, +\,\, \partial^{\,\, \mu}\;\, \phi\;  (\, \mathbf {x}\, ,\; t\, )\,\, ,\,\, \forall\;\, \phi\,\,  
(\, \mathbf {x}\, ,\; t\, )\; ,
\end {equation}
equations (\ref {equazioni di Maxwell per i potenziali}) admit a vast number of solutions; 
all vector potentials that are related to a given solution by a transformation as in 
(\ref {trasformazioni di gauge per i potenziali}) satisfy the same equations and 
are equally suited to describe a given physical configuration.
In particular, equations (\ref {equazioni di Maxwell per i potenziali}) cannot be formulated 
in terms of canonical variables, since a solution can be changed into a different one by 
means of a gauge transformation not affecting the Cauchy data.

The set of solutions can be restricted by means of an auxiliary condition, the procedure being known as the 
choice of a gauge; for such a restriction not to lead to a loss of physical generality, the requirement on the 
constraint is made that a vector potential that fulfills it can be obtained from an arbitrary one, by means 
of a gauge transformation.

An important property is that the solutions cannot be local, because of Gauss' law; this can be easily seen 
for example in the Coulomb gauge.
The starting point for the canonical quantization is the formulation of classical electrodynamics in a local and 
covariant gauge; a gauge of this kind is singled out by the addition of a term to the Maxwell lagrangian, 
which, by weakening the Gauss law, allows for the existence of four-vector potentials obeying covariant 
field equations.
In particular, the Feynman gauge is characterized by the equations of motion
\begin {equation}\label {equazioni in Gupta}
\Box\,\, A^{\,\, \mu}\,\, (\, \mathbf {x}\, ,\; t\, )\, =\,\, j^{\,\, \mu}\; (\, \mathbf {x}\, ,\; t\, )
\end {equation}
and by the subsidiary condition 
\begin {equation}\label {vincolo di Gauss}
\partial^{\;\, \mu}\,\, (\, \partial\, \cdot\, A\, )\; (\, \mathbf {x}\, ,\; t\, )\, =\,\, 0\,\, .\;\;
\end {equation}
The idea is that in principle it might be simpler to solve the hyperbolic dynamics given by 
(\ref {equazioni in Gupta}) and to deal with non locality at a later stage, when imposing 
the Gauss constraint, rather than to have to face directly the dynamical problem 
(\ref {equazioni di Maxwell per i potenziali}), involving non-local fields.
The relevant point to specify is how condition (\ref {vincolo di Gauss}) has to be imposed 
in order to allow to employ local solutions in the construction of potentials obeying 
Maxwell's equations.
A solution of equation (\ref {equazioni in Gupta}) will be called physical if it also satisfies 
(\ref {vincolo di Gauss}).

The strategy at the basis of the Gupta-Bleuler formulation consists in first evaluating local solutions 
of (\ref {equazioni in Gupta}), without constraints, and then in acting on them with suitable 
transformations in order to obtain solutions of (\ref {equazioni di Maxwell per i potenziali}).
Starting from a given $\, A_{\,\, \mu}^{\;\, G}\, $ solving (\ref {equazioni in Gupta}), one can obtain physical 
four-vector potentials $\, A_{\,\, \mu}^{\,\, M}\, $ by means of the transformation
\begin {equation}\label {morfismo Gupta}
A_{\,\, \mu}^{\,\, M}\; (\, \mathbf {x}\, ,\; t\, )\, =\; 
A_{\,\, \mu}^{\;\, G}\; (\, \mathbf {x}\, ,\; t\, )\, +
\; C_{\; \mu\; }^{}(\, \mathbf {x}\, ,\; t\, )\; , 
\quad\quad\quad\quad\quad
\end {equation}
$\, C_{\; \mu}^{}\, $ being any free field whose four-divergence satisfies
\begin {equation}\label {condizione Gupta sulla divergenza}
(\; \partial^{\,\, \mu}\,\,\, C_{\; \mu\, }^{}\, )\, (\, \mathbf {x}\, ,\; t\, )\, =\, -\;
(\; \partial^{\,\, \mu}\,\, A_{\,\, \mu}^{\,\, G}\,\, )\; (\, \mathbf {x}\, ,\; t\, )\; .
\quad\quad\quad\quad\quad
\end {equation}
In classical electrodynamics the behaviour at space-like infinity implied by Maxwell's equations can thus be 
restored with the help of a free field, hence without changing the charge-current distribution.
As discussed before, the support of $\, C\, $ cannot be localized in a bounded space-time region.
The above formulae also show that an assigned configuration of $\, A^{\,\, M}\, $ at space-like infinity, 
described by a field $\, C\, $ obeying (\ref {condizione Gupta sulla divergenza}), is compatible 
with a class of solutions of the wave equation, differing by a divergence-less free field; 
this is related to the fact that the $\, FGB\, $ formulation is invariant with respect to a 
residual group of local gauge transformations.

A convenient procedure to determine a physical solution with given space-like asymptotics is to consider 
$\, A_{\,\, G}^{\,\, \mu\; }= A_{\; (\, j\, )}^{\,\, \mu}\, $ among the solutions of the wave equation.
One can then obtain a vector potential $\, A^{\,\, M}\, $ solving the Maxwell equations for given Cauchy data, by 
means of the transformation (\ref {morfismo Gupta}), with $\, C_{\,\, \mu\, }^{}=\, C_{\,\, \mu\, }^{\;\, (\, j\, )}\, $ 
fulfilling
\begin {equation}\label {condizione Gupta minimale sulla divergenza}
(\; \partial^{\,\, \mu}\;\,\, C_{\,\, \mu\, }^{\,\, (\, j\, )}\; )\; (\, \mathbf {x}\, ,\; t\, )\, =\; -
\; (\; \partial^{\;\, \mu}\,\,\, A_{\,\, \mu}^{\,\, (\, j\, )}\; )\; (\, \mathbf {x}\, ,\; t\, )\,\, .\,
\quad\quad\quad\quad\quad\quad\quad\quad
\end {equation}
The Cauchy data of $\, A^{\,\, M}\, $ determine $\, C\, $ and viceversa:
\begin {eqnarray}
C_{\,\, \mu}^{\,\, (\, j\, )}\; (\, \mathbf {x}\, ,\; t\, =\, 0\, )\, =\,
\, A_{\,\, \mu}^{\,\, M}\; (\, \mathbf {x}\, ,\; t\, =\, 0\, )\,\, ,
\quad\quad\quad\quad\quad\quad\quad\quad\quad\quad\nonumber\\
\\
\dot {C\, }_{\mu}^{\, (\, j\, )}\; (\, \mathbf {x}\, ,\; t\, =\, 0\, )\, =\,
\, \dot {A}_{\,\, \mu}^{\,\, M}\; (\, \mathbf {x}\, ,\; t\, =\, 0\, )\; .\,
\quad\quad\quad\quad\quad\quad\quad\quad\quad\quad\nonumber
\end {eqnarray}
For instance, one gets the Coulomb solution
\begin {equation}
A_{\,\, \mu}^{\;\, (\, Coul.\, )}\,\, (\, \mathbf {x}\, ,\; t\, )\; \equiv
\,\, A_{\,\, \mu}^{\;\, (\, j\, )}\; (\, \mathbf {x}\, ,\; t\, )\, +\; 
C_{\,\, \mu}^{\,\,\, (\, Coul.\, )}\; (\, \mathbf {x}\, ,\; t\, )\,\, ,
\quad\quad\quad\quad
\end {equation}
with $\, C_{\,\, \mu}^{\,\,\, (\, Coul.\, )}\, $ obeying the free wave equation with initial data
\begin {equation}\label {condizioni iniziali coulomb}
C_{\,\, 0}^{\,\,\, (\, Coul.\, )}\,\, (\, \mathbf {x}\, ,\; t\, =\, 0\, )\; \equiv\; \int\,\, d^{\,\, 3\, }y
\;\,\,\, \frac {1} {4\,\, \pi\;\, \vert\, \mathbf {x}\, -\, \mathbf {y}\, \vert}\;\,\, j_{\;\, 0}^{}
\; (\, \mathbf {y}\, ,\; t\, =\, 0\, )\; ,
\end {equation}
\begin {equation}
\dot {C\, }_{\, 0}^{\; (\, Coul.\, )}\,\, (\, \mathbf {x}\, ,\; t\, =\, 0\, )\, \equiv\,\, 0\; ,
\quad\quad\quad\quad\quad\quad\quad\quad\quad\quad\quad\quad\quad\quad\quad\quad\;\, 
\end {equation}
\begin {equation}\label {condizioni Cauchy Coulomb}
C_{\;\, i}^{\,\, (\, Coul.\, )}\; (\, \mathbf {x}\, ,\; t\, =\, 0\, )\, =\;\, 0\; =\,\, 
\dot {C\; }_{i}^{\, (\, Coul.\, )}\; (\, \mathbf {x}\, ,\; t\, =\, 0\, )\; .
\quad\quad\quad\quad\quad\;
\end {equation}
As a matter of fact, equation (\ref {condizioni Cauchy Coulomb}) also implies 
\begin {equation}
\dot {C\, }_{0}^{\; (\, Coul.\, )}\; (\, \mathbf {x}\, ,\; t\, =\, 0\, )\, =\, -\;
(\; \partial^{\;\, \mu}\,\, A_{\,\, \mu}^{\;\, (\, j\, )}\; )\; (\, \mathbf {x}\, ,
\; t\, =\, 0\, )\, =\,\, 0\,\, ,\quad\quad\quad\quad
\end {equation}
while (\ref {condizioni iniziali coulomb}) follows by 
\begin {equation}\label {relazione Coulomb}
(\, \partial_{\;\, 0}^{}\;\, \partial\, \cdot A^{\,\, (\, j\, )}\, )\; (\, \mathbf {x}\, ,\; t\, =\, 0\, )\, =
\; \Box\; A_{\;\, 0}^{\;\, (\, j\, )}\; (\, \mathbf {x}\, ,\; t\, =\, 0\, )\, =
\; j_{\;\, 0}^{}\; (\, \mathbf {x}\, ,\; t\, =\, 0\, )\; .\nonumber
\end {equation}
We can now discuss the features of the solutions of the classical system governed by equations 
(\ref {equazioni in Gupta}), (\ref {vincolo di Gauss}), corresponding to the charge-current 
density
\begin {equation}\label {corrente equazioni Gupta}
j_{\,\, \mathbf {v}}^{\,\, \mu}\; (\, \mathbf {x}\, ,\; t\, )\, =\,\, e\,\,\, v^{\,\, \mu}
\;\,\, \delta\; (\, \mathbf {x}\, -\, \mathbf {v}\,\, t\, )\; .
\quad\quad\quad
\end {equation}
A state of the system is determined by the Cauchy data for the four-vector potential and by the value of the 
velocity of the charge.
The states can be arranged into classes according to their space-like asymptotics; a state is assigned to the 
class $\, \mathscr {C}_{{\; c}_{\; L\, W}^{}}^{},\, $ labeled by a time-like four-vector $\, c_{\, L\, W}^{}\, ,\, $ 
if its initial data (satisfying the Gauss law constraint) can be written as
\begin {eqnarray}\label {condizioni iniziali classi}
A^{\,\, \mu\,\, }(\, \mathbf {x}\, ,\; t\, =\, 0\, )\, =\,\, A_{\,\, c_{\, L\, W\, }^{}}^{\; \mu\;}
(\, \mathbf {x}\, ,\; t\, =\, 0\, )\, +\; g^{\; \mu\; }(\, \mathbf {x}\, ,\; t\, =\, 0\, )
\,\, ,
\;\;\nonumber\\
\\
\dot {A}^{\,\, \mu\;\, }(\, \mathbf {x}\, ,\; t\, =\, 0\, )\, =\,\, 
\dot {A}_{\,\, c_{\, L\, W\, }^{}}^{\; \mu\;\, }(\, \mathbf {x}\; ,
\; t\, =\, 0\, )\, +\; \dot {g\, }^{\, \mu\;\, }(\,\, \mathbf {x}\; ,
\; t\, =\, 0\, )\,\, .\; 
\nonumber
\end {eqnarray}
In (\ref {condizioni iniziali classi}) $\, A_{\,\, c_{\, L\, W\, }^{}}^{\,\, \mu\;\, }(\, \mathbf {x}\, ,\, t=0\, )\, ,\, \dot {A}_{\,\, c_{\, L\, W\, }^{}}^{\,\, \mu\;\, }(\, \mathbf {x}\, ,\, t=0\, )\, $ are the Cauchy data for the Li\'enard-Wiechert solution of equations (\ref {equazioni in Gupta}), (\ref {vincolo di Gauss}), with a constant velocity $\, \mathbf {v}= 
\mathbf {c}_{\; L\, W}^{}\, $ in (\ref {corrente equazioni Gupta}), and the second term on the right-hand side is a contribution with space-like infinity behaviour given by
 \begin {equation}
\partial^{\,\, \mu\;\,\, }g^{\,\, \nu\; }(\, \mathbf {x}\, ,\; t\, =\, 0\, )\, -
\, \partial^{\,\, \nu\;\; }g^{\,\, \mu\; }(\, \mathbf {x}\, ,\; t\,=\, 0\, )\, 
\sim\; o\; (\; \mathbf {x}^{\; -\, 2}\; )\; .
\end {equation}
Each four-vector potential belonging to $\; \mathscr {C}_{\; c_{\, L\, W}^{}}^{}\, $ and solving equations 
(\ref {equazioni in Gupta}), (\ref {vincolo di Gauss}) for a velocity $\, \mathbf {v}\,\, (\; \mathbf {v}
\neq\mathbf {c}_{\; L\, W}^{}\, ) $ can be written as 
\begin {equation}\label {scomposizione potenziali classici}
A^{\,\, \mu\;\, }(\, \mathbf {x}\, ,\; t\, )\, =\,\, A_{\,\, c_{\, L\, W}^{}\; ,\;\, v\;\, }^{\,\, \mu\;\;\, }
(\, \mathbf {x}\, ,\; t\, )\, +\, A_{\,\, v}^{\,\, \mu\;\, }(\, \mathbf {x}\, ,\; t\, )\; ,
\quad\quad\quad\quad\quad\quad
\end {equation}
with $\, A_{\,\, v\;\, }^{\,\, \mu\;}$ the $\, L\, W\, $ solution of eqs.(\ref {equazioni in Gupta}), (\ref {vincolo di Gauss}) 
for a current given by (\ref {corrente equazioni Gupta}) and $\, A_{\,\, c_{\, L\, W}^{}\, ,\;\, v}^{\,\, \mu}\, $ 
a free four-vector potential. 
The corresponding free electromagnetic field 
\begin {equation}
F_{\;\, c_{\, L\, W}^{}\, ,\;\, v}^{\,\, \mu\; \nu}\; (\, \mathbf {x}\, ,\; t\, )\, \equiv\,\, \partial^{\;\, \mu\;\, }
A_{\,\, c_{\, L\, W}^{}\, ,\;\, v\;\, }^{\;\, \nu\;\;\, }(\, \mathbf {x}\, ,\; t\, )\, -\, \partial^{\;\, \nu\;\, }
A_{\,\, c_{\, L\, W}^{}\, ,\;\, v\;\, }^{\;\, \mu\;\;\, }(\, \mathbf {x}\, ,\; t\, )\;\; 
\end {equation} 
satisfies the Maxwell equations, depends both on $\, v\, $ and on $\, c_{\; L\, W}^{}\, $ and for each $\, t\; $ 
has a non-compact support in position space, extending to space-like infinity; precisely, it decays as 
$\, \mathbf {x}^{\; -\, 2}\, .\, $ 
This field is therefore present in all states belonging to the class $\, \mathscr {C}_{\; c_{\, L\, W\, }^{}}^{},\, $ except for $\, \mathbf {v}=\mathbf {c}_{\, L\, W\, }^{}.\, $
Its functional dependence stems from the fact that by definition it must restore the Cauchy data for a 
state in $\, \mathscr {C}_{\; c_{\, L\, W\, }^{}}^{},\, $ since the initial conditions relative to $\, A_{\,\, v}^{\,\, \mu}\, $ 
in (\ref {scomposizione potenziali classici}) belong to a different class.

The classes introduced for the classical system correspond to superselection sectors in the quantum theory. 
In particular, all the states of the class $\, \mathscr {C}_{\; c}^{}\, $ share the same space-like asymptotics, 
governed by the Li\'enard-Wiechert parameter $\, c\, ,\, $ and the representations of the observable algebra 
defined by states belonging to different classes are inequivalent, since the electric flux at space-like 
infinity takes a different value in each class and by locality cannot be modified by the application of 
observables.\\
Furthermore, as we shall see better below, the functional dependence of the classical radiation field 
$\, F_{\;\, c_{\; L\, W}^{}\; ,\;\, v}^{\,\, \mu\; \nu}\, $ corresponds to a non-Fock representation for 
the asymptotic electromagnetic algebra of the quantum system, if $\, \mathbf {v}\, $ is 
interpreted as the asymptotic velocity of the charge.
These arguments have been developed in the lectures on the infrared problem of Ref.~\onlinecite {MoStErice} 
and have also been taken as a guide for non-relativistic $\, QED\, $ in Ref.~\onlinecite {ChFrPiI}. 

We can now take in consideration the quantum case.
In the $\, GB\, $ quantization, physical states are singled out by the auxiliary condition
\begin {eqnarray}\label {condizione di Gupta-Bleuler}
(\, \partial\; \cdot\, A\, )^{\,\, (\, -\, )}\,\,\, \psi\; =\,\, 0\,\, . 
\end {eqnarray}
Since  in the $\, FGB\, $ gauge $\, \partial\, \cdot A\, $ is a free field, its decomposition in negative and 
positive frequencies components is well defined.

For charged states, the existence of solutions of equation (\ref {condizione di Gupta-Bleuler}) in the Gupta-Bleuler 
space $\, \mathscr {G}\, $ of local states, constructed with the aid of the local gauge fields, is excluded by the Gauss law and their 
construction involves a non-local procedure, which is far from trivial, as noted by Zwanziger\cite {Zwanphys}.
Such a construction substantially requires to appropriately implement conditions 
(\ref {morfismo Gupta}), (\ref {condizione Gupta sulla divergenza}).
Actually, since locality of the charged fields is incompatible with positivity, the metric of $\, \mathscr {G}\, $ is indefinite; 
therefore, the construction of physical charged states cannot rely on a completion of $\, \mathscr {G}\, $ based on a 
standard Hilbert closure.

The discussion of main features of the $\, GB\, $ formulation in classical electrodynamics, presented and the 
analysis carried out in the previous Section suggest how to proceed to construct physical charged states in 
the $\, BN\, $ model.

The functionals corresponding to vectors of the space (\ref {sottospazio positivo di Gupta nel modello}), and in 
particular those of the form $\, \omega_{\,\, G}^{}\equiv\, \omega_{\,\, \psi\, }^{}\otimes\, \omega_{\,\, F}^{}\, ,\, $ are positive on $\, \mathscr {A}_{\; obs}^{}\, .\, $  
In order to construct physical charged states, we introduce the automorphism of $\, \mathscr {A}_{\; obs}^{}\, $ defined by
\begin {equation}\label {morfismo sulla particella}
\;\; \tilde {\alpha}\; (\, \hat {\mathbf {x}}\, )\, =\;\, \hat {\mathbf {x}}\;\, , 
\;\, \tilde {\alpha}\; (\, v\, )\, =\;\, v\,\, ,
\end {equation}
\begin {equation}\label {morfismo generale stati fisici}
\; \tilde {\alpha}\; (\, F^{\,\, \mu\; \nu}\; (\, \mathbf {x}\, ,\; t\, )\, )\, =\,\, 
F^{\,\, \mu\; \nu}\; (\, \mathbf {x}\, ,\; t\, )\, +\; e\;\, G^{\,\, \mu\; \nu}\;
(\, \mathbf {x}\, ,\; t\, ;\; \hat {\mathbf {y}}\, )\,\, ,
\end {equation}
\begin {equation}\label {condizione Gauss sul morfismo}
(\; \partial_{\,\, \mu\, }^{}\,\, G^{\,\, \mu\; \nu}\; )\,\, (\, \mathbf {x}\, ,\; t\, ;\; \hat {\mathbf {y}}\, )\, =
\,\, \partial^{\;\, \nu}\,\, (\, \partial\, \cdot\, F\, )\,\, (\, \mathbf {x}\, ,\; t\, ;\; \hat {\mathbf {y}}\, )\; .
\end {equation}
The term on the r.h.s. of (\ref {condizione Gauss sul morfismo}) involves the four-divergence of the 
(operator-valued) four-vector function $\, F_{\,\, v}^{\; \mu}\, $, given by eq.(\ref {funzione F in soluzione Gupta}), and
$\, \hat {\mathbf {y}}\, $ is the position operator of the charged particle.
By equations (\ref {morfismo generale stati fisici}), (\ref {condizione Gauss sul morfismo}) and by the 
definition of $\, \mathcal {K}\, ,\, $ we conclude that the functional $\, \tilde {\omega}_{\,\, \phi}^{}\, ,\, $ 
with expectations given by
\begin {equation}
\tilde {\omega}_{\,\, \phi}^{}\,\, (\, \mathscr {A}_{\; obs}^{}\, )\, \equiv\;\, \omega_{\,\, \phi}^{}\,\, 
(\,\, \tilde {\alpha}\,\, (\, \mathscr {A}_{\; obs}^{}\, )\, )\, =\; \langle\; \phi\,\, ,\; \tilde {\alpha}\; 
(\, \mathscr {A}_{\; obs}^{}\, )\;\, \phi\; \rangle\,\, ,
\end {equation}
is a physical charged state, $\; \forall\; \phi\, \in\, \mathcal {K}\; .$

The classification of such states is most easily done by employing the formulation in terms of 
quantum vector potentials.
The automorphism (\ref {morfismo generale stati fisici}), (\ref {condizione Gauss sul morfismo}) is induced 
by the transformation 
\begin {equation}\label {trasformazione sui potenziali}
\tilde {A}^{\,\, \mu}\; (\, \mathbf {x}\, ,\; t\, )\, \equiv\,\, A^{\,\, \mu}\,\, (\, \mathbf {x}\, ,\; t\, )\, +
\; e\,\,\, C^{\,\, \mu}\,\, (\, \mathbf {x}\, ,\; t\; ;\, \hat {\mathbf {y}}\, )\; ,
\quad\quad\quad\quad
\end {equation}
with $\, C^{\,\, \mu}\, $ a multiplication operator in the particle space, obeying a free dynamics in 
the variables $\, \mathbf {x}\, ,\, t\, $ and with four-divergence fulfilling 
\begin {equation}\label {condizione operatoriale divergenza}
(\, \partial\, \cdot\, C\, )\,\, (\, \mathbf {x}\, ,\; t\: ;\, \hat {\mathbf {y}}\, )\, =\, -
\; (\, \partial\, \cdot\, F\, )\,\, (\, \mathbf {x}\, ,\; t\: ;\, \hat {\mathbf {y}}\, )\; .
\quad\quad\quad\quad
\end {equation}
In analogy with the classical theory, the Coulomb solution can be obtained by solving 
(\ref {condizione operatoriale divergenza}) with the (operatorial) conditions
$\, C_{\,\, i}^{}\: (\: \mathbf {x}\, ,\, t=0\: ;\: \hat {\mathbf {y}}\, )=0=
\dot {C\, }_{i}^{}\: (\: \mathbf {x}\, ,\, t=0\: ;\: \hat {\mathbf {y}}\, )\, .\, $ 
By means of the transformation 
\begin {equation}
\tilde {A}_{\,\, Coul.\; }^{\,\, \mu}(\, \mathbf {x}\, ,\; t\, )\, \equiv\,\, 
A^{\,\, \mu}\; (\, \mathbf {x}\, ,\; t\, )\, +\; C_{\,\, Coul.\; }^{\;\, \mu}
(\; \mathbf {x}\, ,\; t\, ;\; \hat {\mathbf {y}}\; )\; ,
\quad\quad\quad\quad\quad\quad\quad\quad\quad\quad
\end {equation}
with (by also employing eq.(\ref {definizione funzione caratteristica}))
\begin {equation}\label {traslazione Coulomb}
C_{\,\, Coul.}^{\,\, \mu}\, (\, \mathbf {x}\, ,\; t\; ;\, \hat {\mathbf {y}}\; )\, =\,\, \int\,\, d^{\,\, 3\, }z
\;\;\, \rho\,\, (\, \vert\, \mathbf {x}\, -\, \hat {\mathbf {y}}\, -\, \mathbf {z}\, \vert\; )\;\;\, 
\frac {\, \chi_{\,\, \vert\, \mathbf {z}\, \vert\; >\,\, \vert\, t\, \vert}^{}} {4\,\, \pi\,\, \vert\, \mathbf {z}
\, \vert}\;\;\, \delta^{\;\, \mu\,\, 0}\;\, ,\quad\quad\;
\end {equation}
one defines
\begin {equation}\label {stato coulomb modello}
\omega_{\,\, Coul.}^{}\, (\, A^{\; \mu}\, (\, \mathbf {x}\, ,\; t\, )\, )\, \equiv\,\, 
\omega_{\,\, G}^{}\; (\, \tilde {A}_{\,\, Coul.}^{\,\, \mu}\, (\, \mathbf {x}\, ,\; t\, )\, )\: .
\quad\quad\quad\quad\quad\quad\quad\quad\quad\quad\quad\quad
\end {equation}
By employing the covariance of (\ref {condizione operatoriale divergenza}), which holds 
apart from the ultraviolet cutoff, we can construct physical charged states with $\, L\, W\, $ 
space-like asymptotics; with the aid of the transformation
\begin {equation}\label {automorfismo LW}
\tilde {A}_{\; L\, W}^{\,\, \mu}\, (\, \mathbf {x}\, ,\; t\, )\, \equiv\,\, A^{\,\, \mu}\; (\, \mathbf {x}\, ,\; t\, )\, +
\; C_{\,\, c_{\: L\, W}^{}}^{\,\, \mu}\, (\, \mathbf {x}\, ,\; t\, ;\; \hat {\mathbf {y}}\, )\; ,
\quad\quad\quad\quad\quad\quad\quad\quad\quad\quad\quad\quad\quad
\end {equation}
\begin {align}\label {termine aggiunto LW}
C_{\,\, c_{\: L\, W}^{}}^{\,\, \mu}\, (\, \mathbf {x}\, ,\; t\; ;\, \hat {\mathbf {y}}\, )\, =
\,\, \int\,\, d^{\,\, 4\, }z\,\,\,\, \rho\,\, (\, \vert\, \mathbf {x}\, -\, \hat {\mathbf {y}}\, -
\, \underline {\Lambda_{\; c_{\, L\, W}^{}}^{}\; z}\, \vert\, )\;\,\, \,
\delta\; (\, (\; \Lambda_{\,\, c_{\, L\, W}^{}}^{}\; z\; )_{\;\, 0}^{}\; )
\quad\quad\quad
\nonumber\\
\times\;\, \frac {{c_{\, L\, W}^{\;\, \mu}}\,\,\,\, \chi_{\; z^{\, 2}\; <\,\, 0}^{}} 
{4\,\, \pi\,\,\, [\; (\, c_{\, L\, W}^{}\cdot\, z\, )^{\,\, 2}\, -\, 
c_{\; L\, W}^{\,\, 2}\,\, z^{\,\, 2}\;\, ]^{\,\, 1\, /\; 2}\, }\;\, ,
\end {align}
one obtains physical charged states labeled by the time-like Li\'enard-Wiechert four-vector $\, c_{\; L\, W}^{}:$
\begin {equation}\label {stato LW modello}
\omega_{\; L\, W}^{}\; (\, A^{\,\, \mu\; }(\, \mathbf {x}\, ,\: t\, )\, )\, \equiv
\,\, \omega_{\,\, G}^{}\; (\, \tilde {A}_{\, L\, W}^{\,\, \mu}\,
(\, \mathbf {x}\, ,\; t\, )\, )\; .\quad\quad\quad\quad\quad\quad\quad\quad\quad\quad\quad\quad
\end {equation}
Such states, which will be referred to in the sequel as Li\'enard-Wiechert states, are constructed via an 
automorphism which amounts to a shift of the contribution from the hyperbolic dynamics by a non-local 
solution of the free wave equation, in analogy with the classical theory. 
The field (\ref {termine aggiunto LW}) obeys indeed a free evolution, because it is the convolution with 
(the Lorentz-transformed of) a (Coulomb) field obeying a free wave equation.

It is interesting to point out that, although they are not obtained from vectors of $\, \mathscr {G}\, $ through a limiting 
procedure, the states (\ref {stato LW modello}) are nevertheless quite linked to the indefinite-metric 
space, since their expectations on the subalgebras $\, \mathscr {A}\, (\, \mathcal {O}_{\; +}^{}\, )\, $ and $\, \mathscr {A}\, 
(\, \mathcal {O}_{\; -}^{}\, )\, $ equal those 
of $\, \omega_{\,\, G}^{}\, .\, $
Such a property of the $\, L\, W\, $ states ultimately relies on the existence of solutions of the free wave equation 
with support in $\, \mathcal {O}\; '.\, $

Now we wish to analyze how states with $\, L\, W\, $ asymptotics can be constructed by means of a procedure 
employing the asymptotic electromagnetic fields.
Let us introduce the operators (for definiteness, we consider the $\, out\, $ vector potential, the 
treatment for the $\, in\, $ field being analogous, with obvious changes)
\begin {eqnarray}\label {operatori unitari regolarizzati}
U_{\, R\;\, }^{}(\, C\, )\, \equiv\,\, \exp\,\, (\, -\; i\,\, e\, \int\,\, d^{\,\, 3\, }x
\,\,\, A_{\; out}^{\; \mu}\; (\, \mathbf {x}\, ,\: t\, )\;\, \overleftrightarrow 
{\;\, \partial_{\;\, t}^{}}\;\,\, C_{\, R\; ,\,\, \mu}^{}\, (\: \mathbf {x\, }-\, 
\hat {\mathbf {y}}\, ,\: t\, )\, =\;\;\;\;\;\;
\quad\, \nonumber\\
=\; \lim_{t\; \rightarrow\; +\; \infty}\,\, \exp\,\, (\, -\; i\,\, e\, \int\,\, d^{\,\, 3\, }x\,\,\, A^{\; \mu\; }(\, \mathbf {x}\, ,\; t\, )
\;\, \overleftrightarrow {\;\, \partial_{\;\, t}^{}}\;\,\, C_{\, R\; ,\,\, \mu}^{}\, (\: \mathbf {x\, }-\, \hat {\mathbf {y}}\, ,
\; t\, )\; ,\quad\;
\end {eqnarray}
where $\, C_{\, R}^{}\, $ is obtained by regularizing the field $\, C\, ,\, $ introduced in 
(\ref {trasformazione sui potenziali}), (\ref {condizione operatoriale divergenza}) and 
acting as a multiplication operator in the charged particle space, 
as follows
\begin {equation}\label {funzioni di coerenza regolarizzate}
C_{\, R}^{\,\, \mu}\; (\: \mathbf {x}\, -\, \hat {\mathbf {y}}\, ,\; t\, )\, \equiv\,\, 
C^{\; \mu}\: (\: \mathbf {x}\, -\, \hat {\mathbf {y}}\, ,\; t\, )\;\;\, \chi_{\,\, R}^{}
\; (\, \mathbf {x}\, -\, \hat {\mathbf {y}}\, )\; ,\quad\quad\quad\quad
\end {equation}
\begin {equation}\label {regolarizzazione esplicita}
\chi_{\; R}^{}\; (\, \mathbf {x}\, )\, \equiv\,\, \chi\; (\, \frac {\vert\, \mathbf {x}\, \vert} {R\, }\, )\; ,
\; R\, >\, 0\,\, ,\quad\quad\quad\quad\quad\quad
\end {equation}
with $\, \chi\, $ a smooth function with compact support.
The functions (\ref {regolarizzazione esplicita}) serve to cope with the infrared 
divergences of the smeared field caused by the long-range Coulomb tail of the test functions, which must 
necessarily be present for the Gauss law to be fulfilled. 

With the aid of those operators, one can introduce a one-parameter group of automorphisms of 
$\, \mathscr {F}^{\; out}\, ,$
\begin {equation}\label {automorfismo dell'algebra delle osservabili}
\alpha_{\; R}^{}\; (\, D\, )\, \equiv\;\, U_{\; R}^{\,\, -\, 1}\,\, (\, C\, )\;\;\, D
\;\;\, U_{\; R}^{}\,\, (\, C\, )\,\, ,\; D\, \in\, \mathscr {F}^{\,\, out}\,\, ,
\quad\quad\quad\quad\quad\quad\quad\quad
\end {equation}
induced by the transformation 
\begin {equation}\label {espressioni esplicite per gli shift}
U_{\; R}^{\,\, -\, 1}\, (\, C\, )\;\,\, A_{\; out}^{\,\, \mu}\, (\, \mathbf {x}\, ,\; t\, )\;\;
\, U_{\; R}^{}\; (\, C\, )\, =\, A_{\; out}^{\,\, \mu}\, (\, \mathbf {x}\, ,\; t\, )\; +\; 
e\,\,\, C_{\; R}^{\;\, \mu}\; (\, \mathbf {x}\, -\, \hat {\mathbf {y}}\, ,\; t\, )\; ,
\quad\quad
\end {equation}
and by the corresponding one on the time-derivative of the four-vector potential.
In particular, one has
\begin {equation}\label {espressioni esplicite per lo shift su B}
\alpha_{\: R}^{}\; (\, (\, \partial\, \cdot\, A\, )\; (\, \mathbf {x}\, ,\; t\, )\, )\; =
\; (\, \partial\, \cdot\, A\, )\; (\, \mathbf {x}\, ,\; t\, )\, +\; e\;\, 
(\, \partial\, \cdot\, C_{\: R}^{}\, )\; (\, \mathbf {x}\, ,\; t\, )\; .
\quad\quad\quad\quad
\end {equation}
The functionals defined on $\, \mathscr {F}^{\,\, out}\, $ as
\begin {equation}
\omega_{\,\, \psi\; ,\,\, R}^{}\; (\, D\, )\, \equiv\,\, \langle\,\, U_{\; R}^{}\;\, \psi\; ,
\; D\,\,\, U_{\; R}^{}\;\, \psi\,\, \rangle\; ,\,\, \psi\, \in\; \mathcal {K}\; ,
\,\, D\, \in\, \mathscr {F}^{\;\, out}\,\, ,
\quad\quad\quad\quad\quad\quad\quad
\end {equation}
are positive, since
\begin {eqnarray}\label {funzionali positivi a R fisso}
\omega_{\,\, \psi\; ,\,\, R}^{}\; (\, D^{\,\, *}\; D\, )\, =\; \langle\,\, U_{\; R}^{}
\,\,\, \psi\,\, ,\; D^{\,\, *}\; D\,\,\, U_{\; R}^{}\,\,\, \psi\,\, \rangle\, =\; \langle\,\, 
\psi\,\, ,\,\, \alpha_{\,\, R}^{}\; (\, D^{\; *}\; D\, )\,\,\, \psi\,\, \rangle
\quad\quad\quad\quad\; \nonumber\\
=\,\, \omega_{\,\, \psi}^{}\; (\, \alpha_{\; R}^{}\; (\, D^{\,\, *}
\; D\,\, )\, )\, \geq\; 0\,\, .\quad\quad\quad\nonumber 
\end {eqnarray}

Concerning the removal of the infrared cutoff, one needs to specify a notion of convergence.
Let $\, \alpha\, (\, D\, )\equiv\, \lim_{\,\, R\,\, \rightarrow\, +\, \infty}\, \alpha_{\; R}^{}\; (\, D\, )\, ;\, $
the infrared cutoff can then be removed in the expectations of asymptotic observables, 
yielding
\begin {equation}
\omega_{\,\, \psi\,\, }^{}(\, D\, )\; \equiv\, \lim_{\,\, R\,\, \rightarrow\; +\; \infty\; }
\,\, \omega_{\,\, \psi\; ,\,\, R\;\, }^{}(\, D\, )\, =\;\, \omega_{\,\, \psi\,\, }^{}
(\, \alpha\, (\, D\, )\, )\; .\,\,
\end {equation}
The limiting functionals are positive, 
\begin {equation}
\omega_{\,\, \psi\;\, }^{}(\, D^{\; *\,\, }D\, )\; =\lim_{\,\, R\,\, \rightarrow\; +\; \infty}
\,\, \omega_{\,\, \psi\; ,\,\, R\;\, }^{}(\, D^{\; *\,\, }D\, )\, \geq\; 0\,\, ,\quad\quad
\end {equation}
and are physical states, since, by (\ref {espressioni esplicite per lo shift su B}),
(\ref {condizione operatoriale divergenza}) and the definition of 
$\, \mathcal {K}\, $, 
\begin {equation}
\omega_{\,\, \psi}^{}\; (\, (\, \partial\, \cdot\, A\, )\; (\, \mathbf {x}\, ,\; t\, )\, )\; =
\lim_{\,\, R\; \rightarrow\; +\; \infty\; }\, \omega_{\,\, \psi}^{}\,\, (\,\, \alpha_{\: R}^{}\; 
(\, (\, \partial\, \cdot\, A\, )\; (\, \mathbf {x}\, ,\; t\, )\, )\, )\, =\,\, 0\; .
\end {equation}

A relevant property of Dirac-type factors as (\ref {operatori unitari regolarizzati}) is that by Huyghens' principle
they preserve the expectations of e.m. observables within forward and backward lightcones, since they are 
obtained as (suitably regularized) time limits of the interacting four-vector potential, smeared with 
solutions of the free wave equation.
It would be of interest to investigate whether the use of such exponentials, which can be regarded as alternative 
with respect to the approach based on the construction of Dirac-type exponentials by means of interacting 
gauge fields\cite {Steinb,BDMRS01}, may be successfully extended to $\, QED\, $.

In the sequel we show that the $\, L\, W\, $ charged states, which belong to superselection sectors labeled 
by different values of $\, c_{\; L\, W}^{}\, ,\, $ can be arranged in a charge class, a concept introduced by Buchholz
\cite {Buch82} in a general analysis of the state space of Quantum Electrodynamics.  

We recall that positive energy representations of the observable algebra, with given electric charge, factorial in 
a forward lightcone $\, V_{\,\, +}^{}\, $ and possibly belonging to different superselection sectors, are assigned 
to the same charge class if their restrictions to $\, V_{\,\, +}^{}\, $ are equivalent. 

This concept is physically motivated by the kinematic consideration that since electrically charged particle are 
massive they have to eventually enter any forward lightcone and thus it should be possible to determine the 
total charge of a state by measurements performed in such a region.
In this way one should be able to distinguish the electric charge among the superselection rules in $\, QED\, ,\, $ 
since, due to the possible presence of photons coming from asymptotic negative times, measurements in a 
forward lightcone are not enough to determine the value of the electric flux at space-like infinity in a given 
representation; each charge class should then contain superselection sectors with a given electric charge 
but different flux-distributions.

Buchholz showed that given an irreducible positive energy representation $\, \pi\, $ of the observable algebra in a 
Hilbert space $\, \mathscr {H}\, ,\, $ the restrictions to the subalgebras $\, \mathscr {F}^{\; out\; }(\, V_{\,\, +}^{}\, )\, $ of the representations belonging to $\, [\, \pi\, ]\, $ are 
equivalent; therefore, the sectors in a charge class cannot be distinguished by measurements of the outgoing 
electromagnetic fields in the forward lightcone.

He also proved that by adding an arbitrary number of low-energy photons to a given 
$\, \Psi\in\mathscr {H}\, $ 
one can construct a state, with finite energy and the same charge, inducing a 
representation of $\, \mathscr {F}^{\; out\; }(\, V_{\,\, +}^{}\, )\, $ inequivalent 
to $\, \pi\, (\, \mathscr {F}^{\: out\; }(\, V_{\,\, +}^{}\, )\, )\, $ 
and thus corresponding to a different charge class.
Therefore, charge classes are in general characterized not only by the total electric 
charge of their states, but also by the presence of background radiation fields. 
A criterion of infrared minimality can be introduced, by demanding that it selects the 
charge classes whose superselection sectors have the best possible localization 
properties with respect to the vacuum, that is, they do not have any background 
radiation field. Such sectors are expected to be a convenient set for a systematic 
analysis of the infrared problem.

We shall now discuss the space-time properties of the $\, L\, \, W\, $ physical charged states of the model.
First, such states coincide on $\, \mathscr {A}\, (\, \mathcal {O}_{\; +}^{}\, )\, $, since the free field 
$\, C^{\,\, \mu}\, $ has support in $\, \mathcal {O\; }',\, $ and are given by $\, \omega_{\; G}^{}\, $, which 
is positive and satisfies the auxiliary condition in $\, \mathcal {O}_{\; +\, }^{};\, $ product functionals defined
within the $\, GB\, $ formulation identify therefore a \emph {unique} charge class.
Secondly, equations  (\ref {automorfismo LW}), (\ref {termine aggiunto LW}) imply that the
$\, L\, W\, $ states induce Fock representations of the subalgebras $\, \mathscr {F}^{\; out\,\, }
(\, \mathcal {O}_{\; +}^{}\, )\, $ and $\, \mathscr {F}^{\; out\,\, }
(\, \mathcal {O}_{\, -}^{}\, )\, ;\, $ 
this gives the simplest result for the indetermination left by Buchholz's treatment for 
the representations of $\, \mathscr {F}^{\; as\,\, } 
(\, \mathcal {O}_{\; +}^{}\, )\, $ and $\, \mathscr {F}^{\; as\,\, }
(\, \mathcal {O}_{\, -}^{}\, )\; .$

The Gupta-Bleuler charge class can thus be identified as the charge class containing all representations 
$\, \pi\, $ of the algebra of observables whose restrictions to $\, \mathscr {F}^{\; out\,\, } (\, \mathcal {O}_{\; +}^{}\, )\, $ and $\, \mathscr {F}^{\; out\,\, }(\, \mathcal {O}_{\, -}^{}\, )\, $ are Fock.
Since this result can be seen as a consequence of the locality of the gauge fields and of the support 
properties of the automorphism on the asymptotic electromagnetic algebras, one expects that it 
might also hold in the corresponding formulation of $\, QED\, .\, $

It is worthwhile to remark that the condition on $\, \mathscr {F}^{\; out\,\, }(\, \mathcal {O}_{\; +}^{}\, )\, $ agrees with the above-mentioned result that 
the representations of the outgoing electromagnetic field in a charge class cannot be distinguished by 
measurements in the forward lightcone. 
The Fock property of $\, \pi\, (\, \mathscr {F}^{\; out\; }(\, \mathcal {O}_{\; +}^{}\, )\, )\, $ also implies that 
the $\, GB\, $ charge class is infrared minimal in the sense described above, namely, it does not 
contain background radiation fields but only the field associated to the asymptotic momentum of 
the charge.

We also point out that the condition on $\, \mathscr {F}^{\; out\,\, }(\, \mathcal {O}_{\; -}^{}\, )\, $ is due to the fact that the $\, L\, W\, $ states have been 
constructed with the aid of a free field with expectations having support in $\, \mathcal {O}\: ',\, $ and therefore enjoy at 
least some of the locality properties holding in models of field theories with massless bosons and 
standard charges.

We conclude the paper with a discussion concerning the question of how the model should be 
improved in order to account for a proper description of electrically charged particle at large 
times and of fermion-loop effects.

Concerning the first issue, the algebraic theory provides the following general framework.
The elements of the algebra $\, \mathscr {M}^{\; out}\, $ of the observables describing the (charged) massive 
particles at positive asymptotic times (and therefore associated to the intersection of all forward lightcones) 
are compatible with the observables of $\, \mathscr {F}^{\; out}\, ,\, $ owing to Huyghens' principle. 
Such a property should hold in particular for the asymptotic four-velocity of the charge, which it is expected to 
belong to the center of $\, \mathscr {F}^{\; out}\, ,\, $ namely to index inequivalent representations of 
this algebra, as it happens in the $\, BN\, $ model.\\
Under this assumption, the representation of the algebra of all outgoing observables $\, \mathscr {A}^{\; out}\, $ 
can be reduced with respect to the asymptotic momenta of the charges and cannot therefore provide a complete 
characterization of the physical system.
This is related to the fact that an observable describing the asymptotic position of a particle carrying an electric 
charge cannot exist, if it is requested to eventually enter any forward lightcone; in fact, by kinematical reasons
such an observable would necessarily belong to $\, \mathscr {F}^{\; out}\, '\, $ and thereby commute with the 
corresponding asymptotic momentum.

Buchholz suggested that a collision theory employing a complete set of variables for the outgoing charged 
particles may be achieved if one considers observables contained in a \emph {fixed} forward lightcone.
These considerations agree with the conjecture that it may be possible to completely determine a scattering 
matrix by considering a modified asymptotic dynamics for the charges, allowing for the construction of an 
asymptotic position variable, along the lines of Dollard's treatment of Coulomb scattering\cite {Doll64}. 
In fact, the dynamics of the position variable at large (positive) times has to account for the effects of the 
interaction with photons, which are not described by the observables belonging to the intersection of all 
(forward) lightcones.

Concerning the fermion-loop effects, a general theorem states\cite {BDMRS01} that for any subalgebra $\, C\, $ of 
local observables, stable under translations and irreducible in the vacuum sector, the quantum corrections 
cannot vanish for all elements of $\, C\, ;\, $ in particular, it seems to indicate that, as consequence of the 
delocalization caused by vacuum polarization effects, there cannot exist electrically charged states which 
are local with respect to the charge-current density. 

As discussed before, in the $\, BN\, $ model the behaviour at space-like infinity implied by Maxwell's equations 
can be restored with the help of a free field, hence without changing the charge-current distribution; the 
so-obtained $\, L\, W\, $ states are thus local with respect to the charge-current density. 
Of course, this property is not in contradiction with the above-mentioned theorem, since the model does not 
account for loops of fermions, but the point is that the same feature seems to be necessarily implied by 
locality also in $\, QED\, ,\, $ for physical charged states constructed within the $\, GB\, $ formulation
\cite {MoSt2}.
However, since well-known results from the standard diagrammatic expansion imply that contributions from 
fermion loops should vanish for asymptotic times, the relevance of the theorem for electrically charged 
states constructed with the aid of asymptotic fields is unclear. 
This problem certainly deserves further study.

\begin {acknowledgments}
\emph {This paper is based in part on preliminary investigations by G. Morchio and F. Strocchi on the
space-time properties of solvable hamiltonian models, devoted to a better understanding of the 
Gupta-Bleuler formulation of Quantum Electrodynamics. 
I would like to thank G. Morchio for extensive discussions on these topics.}
\end {acknowledgments}

\end {document}